\documentclass{egpubl}

\usepackage[T1]{fontenc}
\usepackage{dfadobe}

\usepackage{cite}
\usepackage{xcolor}
\usepackage{subcaption}
\usepackage{adjustbox}
\usepackage{multirow}
\usepackage{booktabs}
\usepackage{bm}
\usepackage{amsmath}
\usepackage{amssymb}
\usepackage{array}
\usepackage{float}
\usepackage{url}
\usepackage{etoolbox}
\usepackage{eso-pic}

\newcommand{\ygt}{y_\text{GT}}

\newcommand{\yp}{y_\theta}

\newcommand{\Loss}{\mathcal{L}}
\newcommand{\net}{\phi_\theta}
\newcommand{\neta}{\phi_{\theta,\omega}}

\BibtexOrBiblatex

\electronicVersion
\PrintedOrElectronic
\hypersetup{
  pdfsubject={Reference-free evaluation of camera ISP pipelines},
  pdfkeywords={camera ISP, image quality assessment, reference-free evaluation}
}
\usepackage{egweblnk}

\makeatletter
\providecommand{\p@EGlocalpagenumber}{}
\def\ps@titlepage{%
  \let\@mkboth\@gobbletwo
  \def\@oddhead{}%
  \def\@evenhead{}%
  \def\@oddfoot{}%
  \def\@evenfoot{}%
  \let\sectionmark\@gobble
  \let\subsectionmark\@gobble
}

\copyrightTextTitPag{}
\copyrightTextRunPag{}

\makeatother
\AtBeginDocument{%
  \hypersetup{
    pdfsubject={},
    pdfkeywords={}
  }%
}

\title[A Reference-Free Framework for Evaluating Single-Frame ISP Pipelines]%
      {A Reference-Free Framework for \\ Evaluating Single-Frame ISP Pipelines}

\author[Y. Cho et al.]
{
\parbox{\textwidth}{\centering
Yujin Cho$^{1,2}$,
Sira Ferradans$^{2}$,
Jean-Michel Morel$^{3}$,
Gabriele Facciolo$^{1,5}$,
and Thomas Eboli$^{4}$
\\[0.5ex]
{\small
$^{1}$ENS Paris-Saclay, France
\quad
$^{2}$DXOMARK, France
\quad
$^{3}$Lingnan University, Hong Kong
\\[-0.2ex]
$^{4}$CFM, France
\quad
$^{5}$Institut Universitaire de France, France
}
}
}

\begin{document}
\teaser{
  \centering
  \includegraphics[width=0.9\linewidth]{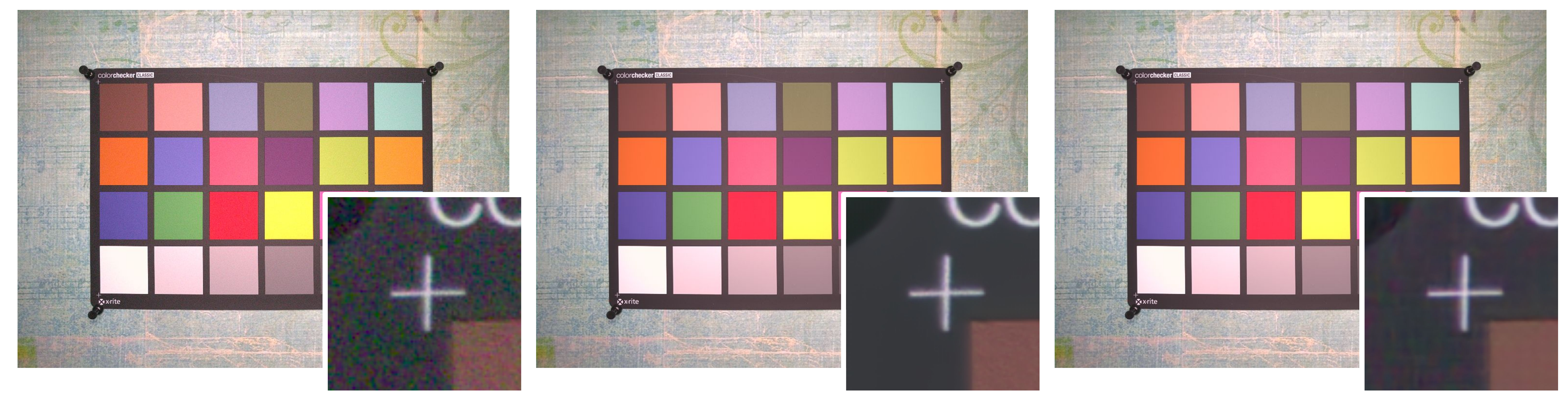}

  \vspace{0.2em} 
  \begin{minipage}[t]{0.3\linewidth}
    \centering Noisy Raw Image \\
    (converted to sRGB for visualization)
  \end{minipage}%
  \begin{minipage}[t]{0.3\linewidth}
    \centering ISP-Processed Image \\
    \textbf{GT PSNR = 35.0dB} \\
    \small {(*not available in practice)}
  \end{minipage}%
  \begin{minipage}[t]{0.3\linewidth}
    \centering Proxy Reference \\
    \textbf{Predicted PSNR = 35.2dB}
  \end{minipage}

  \caption{\textbf{Reference-free metric estimation for camera pipeline evaluation.} A noisy RAW image is first processed by the ISP to produce an sRGB camera output. At inference time, the proposed method takes only this processed image and its ISO metadata as input, predicts a proxy reference, and estimates full-reference metrics without access to the ground-truth reference.
  We used a raw photograph
taken with a Samsung Galaxy S6 Edge phone from~\cite{abdelhamed2018dataset} at ISO 800
that comes with a target image to compute the ground-truth PSNR score for comparison.}
  \label{fig:teaser}
}

\maketitle

\begin{abstract}
Evaluating camera image signal processing (ISP) pipelines requires
measuring low-level artifacts introduced by operations such as denoising,
demosaicing, tone mapping, and compression. Blind image quality assessment
(IQA) techniques can grade visual quality without a reference, but they
typically focus on semantic and high-level visual cues or human perceptual
scores rather than the low-level image-processing artifacts introduced by
camera pipelines. In contrast, full-reference metrics such as PSNR and
SSIM measure pixel-level differences and structural similarity, while
LPIPS measures perceptual similarity in deep feature space. However, these
metrics require perfectly aligned image pairs, which are difficult to
collect in practical settings. We propose a reference-free learning framework that estimates
full-reference image quality metrics from a processed sRGB image and its
ISO metadata. Our method predicts a proxy sRGB reference, which is then
compared with the processed image to compute PSNR, SSIM, and LPIPS in their
standard full-reference form. Our experiments show that the proxy-reference
model can be learned from synthetic data and applied to real camera data.
We further show that lightweight LoRA fine-tuning enables efficient
adaptation when ISP components or pipeline configurations are changed.
The proposed method outperforms direct metric regression in estimating
metric values and achieves higher agreement with full-reference rankings
than conventional blind IQA methods. These results demonstrate the
feasibility of reference-free estimation of full-reference metrics for
practical camera-pipeline evaluation.


\end{abstract}  
\section{Introduction}

With the growth of smartphone photography, millions of images are captured and processed every day, making camera quality evaluation increasingly important. The limited physical space available in smartphones has led manufacturers toward miniaturized sensors with relatively poor signal-to-noise ratios. To compensate for sensor noise and other acquisition limitations, raw images are processed by increasingly complex image signal processing (ISP) pipelines. Given the complexity of current ISPs, the use of image quality assessment (IQA) methods has become an important component in the development of camera pipelines  to perform ISP tuning\cite{tseng2019proxy}. IQA methods are generally divided into reference-based and blind approaches. Full-reference (FR) metrics such as PSNR, SSIM~\cite{wang2004ssim}, HDR-VDP/HDR-VDP-2, and CameraVDP~\cite{mantiuk2005predicting,mantiuk2011hdrvdp,cai2025cameravdp}, and LPIPS~\cite{zhang2018unreasonable} quantify pixel-level, structural, or perceptual differences between a processed image and a pristine reference. Their main limitation is the need for perfectly aligned image pairs acquired under matching conditions. In practice, reproducing exposure, illumination, scene motion, and other acquisition conditions makes such references difficult to obtain.

Blind image quality assessment methods, on the other hand, evaluate images withouot a pristine reference image either by using predefined image statistics, such as NIQE~\cite{mittal2012no}, or models trained on perceptually annotated datasets, such as MUSIQ~\cite{ke2021musiq} and CLIP-IQA~\cite{wang2022exploring}. Many blind IQA methods rely on mean opinion scores (MOS) collected from human raters. However, collecting MOS annotations is costly, and their reliability is affected by inter-annotator variability. As observed for UNIQUE~\cite{zhang2021uncertainty}, this variability can obscure low-level artifacts and reduce consistency across datasets and tasks. Moreover, methods trained to predict subjective quality often emphasize semantic content, aesthetics, and high-level visual cues rather than subtle artifacts produced by ISP pipelines, such as color shifts, blocking, demosaicing errors, and denoising artifacts manifest at the pixel level.
Our goal is different from conventional perceptual blind IQA. 
We consider the evaluation of ISP components and configurations for a target camera with a calibrated sensor noise profile. Within this setting, we aim to estimate structural and perceptual metrics such as PSNR, SSIM and LPIPS without requiring the corresponding clean reference. We do not address universal transfer across different camera sensors or manufacturers. Instead, we focus on low-level artifacts produced when processing images with a given ISP or when modifying individual components of that ISP.

We propose a reference-free framework that transforms these full-reference metrics (PSNR, SSIM, and LPIPS) into learnable blind criteria, without requiring aligned pairs or human annotations. Our approach takes as input a processed image together with its ISO metadata and leverages a restoration network to generate proxy reference. 
With this proxy reference and the input, we can compute the standard metrics in a blind setting. 
Our experiments show that a single model can optimize PSNR and SSIM jointly, while still achieving respectable correlation with LPIPS. This joint formulation makes the framework more practical, since one network can generalize across multiple evaluation criteria without the need for metric-specific variants.
The proposed framework is trained entirely on a fixed synthetic pipeline and generalizes to previously unseen, realistic ISP pipeline such as Adobe Lightroom RAW-to-sRGB. Our evaluation extends beyond denoising, tone mapping, JPEG compression. 
 We also study changes to individual ISP components and configurations, including denoisers, demosaicing method, compression settings, and optical degradation. As such component changes may introduce a domain gap relative to the base training pipeline, we use lightweight LoRA fine-tuning~\cite{hu2022lora} for efficient component- and pipeline-specific adaptation.
  Our contributions are as follows:

\begin{itemize}
    \item We propose a reference-free IQA framework that takes a processed
    image and its ISO metadata as input and estimates structural and perceptual
    metrics, including PSNR, SSIM, and LPIPS, without requiring aligned pairs
    or subjective MOS annotations.

    \item We introduce a multi-metric joint optimization strategy that allows a single model to generalize across diverse stages of the ISP pipeline, including non-linear operations such as tone mapping and JPEG compression.

    \item We demonstrate the robust generalization of our model from synthetic data to unseen ISP pipelines. Furthermore, we show that few-shot LoRA fine-tuning effectively mitigates the domain gap, enabling efficient adaptation to diverse new pipelines with minimal data.
\end{itemize}


 \begin{figure*}
     \centering
     \includegraphics[width=0.98\linewidth]{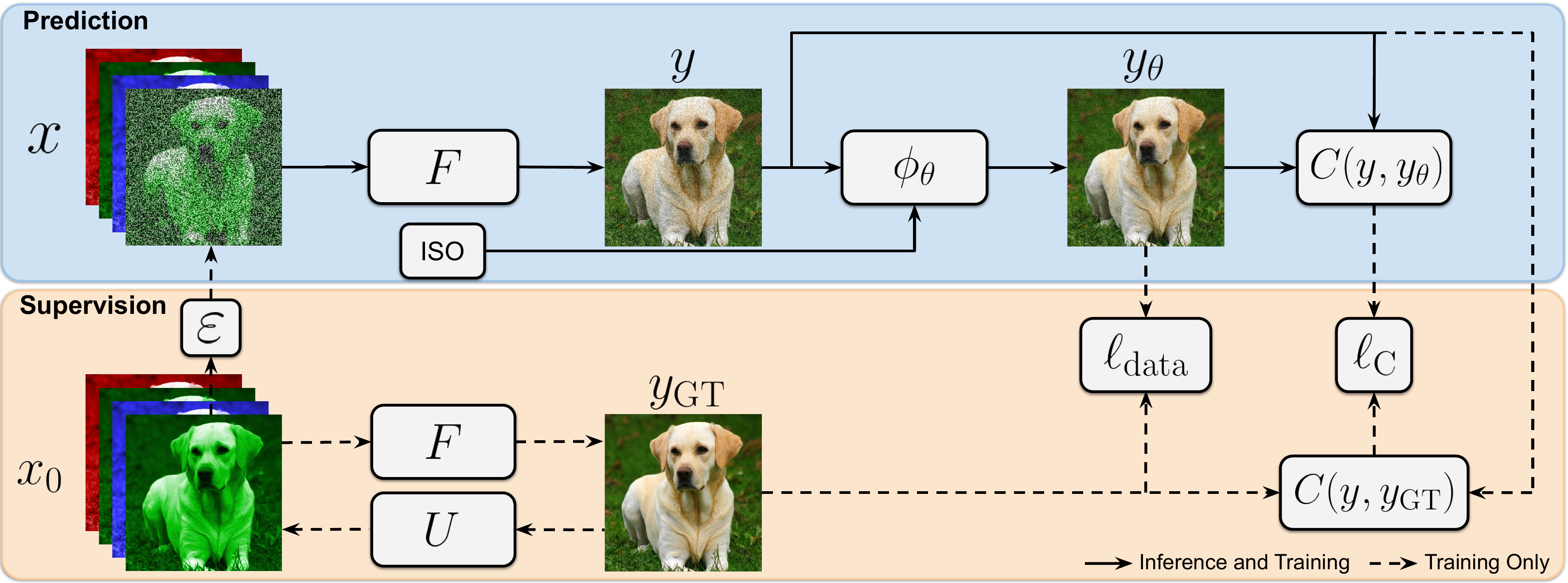}
    \caption{\textbf{Explanatory diagram of our {\em restoration} method.} 
    Top: Main branch of our approach. 
    A noisy raw image $x$ is processed with the pipeline $F$ to obtain the image $y=F(x)$. 
    The image $y$, together with its ISO value, is fed to the network
    $\net$ to predict the proxy reference
    $y_\theta=\net(y,\text{ISO})$ and the score $C(y,y_\theta)$.
    Bottom: Additional branch for the training and evaluation stages. 
    Starting either from a high-quality raw image $x_0$ such as those taken from~\cite{abdelhamed2018dataset} or from a pseudo-raw image obtained by unprocessing a high-quality sRGB reference~\cite{brooks2019unprocessing}, we (1) add noise $\varepsilon$ to generate the input $x$, and (2) generate the target reference $\ygt$ with the pipeline $F$, necessary to compute the ground-truth score $C(y,\ygt)$ and the losses $\ell_\text{data}$ and $\ell_C$ for updating $\theta$.
The perceptual feature loss is omitted from the diagram for clarity.}
     \label{fig:diagram}    
 \end{figure*}

\section{Related Work}

\subsection{Blind evaluation of image denoising}
Stein’s unbiased risk estimate (SURE) was one of the first 
``blind'' estimators of Gaussian noise, offering a way to predict 
mean-squared error without access to a reference image and 
historically used to set denoising 
thresholds~\cite{stein1981estimation}. Since then, only a
handful of blind noise metrics have appeared, mostly designed
and validated for Gaussian noise. The authors of \cite{rakhshanfar2018sparsity}
proposed a sparsity-based index that analyzes patch orientations
and singular-value spectra to estimate residual noise, while the method in
 \cite{lu2020no} combined noise-residual statistics with 
hand-crafted features in a random-forest regressor to rank denoised outputs. Although these approaches work well on classic noise types (Gaussian, Poisson, salt-and-pepper), 
they break down on more complex, signal-dependent 
distributions~\cite{foi2008practical, wei2020physics}.
In contrast, our work predicts full-reference quality metrics like PSNR and
SSIM directly from a single distorted image without any ground-truth 
reference. Our method is flexible enough to adapt to
any realistic noise model of raw photographs.

\subsection{Blind IQA}
Blind image quality assessment proposes metrics consistent with human 
perceptual judgments, usually quantified 
through mean opinion scores (MOS) from perceptual annotations. Initial regression-based 
blind IQA approaches rely on hand-crafted features that captured natural 
scene statistics and image-specific 
characteristics~\cite{moorthy2011blind, mittal2012no, saad2012blind, ye2012unsupervised, mittal2012making} which
may be too coarse to detect the most subtle artifacts.
Recent techniques based on deep learning comprise supervised regressors~\cite{bosse2017deep, su2020blindly, zhang2020blind, ke2021musiq, golestaneh2022no,yang2022maniqa} trained on annotated data,
and self-supervised models~\cite{madhusudana2021st, chen2023human, ghildyal2024trends} pre-trained on large corpora of images.
Most of the training data come from MOS-labeled datasets featuring
image degradations such as severe JPEG compression or noise~\cite{lin2019kadid, lin2020deepfl, ponomarenko2013new} or perceptual quality~\cite{ke2021musiq}.
Conversely, ranking‐based blind IQA methods use pairwise or list-wise losses to learn relative quality orders, improving sensitivity to subtle degradations~\cite{liu2017rankiqa, zhang2021uncertainty}. However, they require carefully curated image pairs and produce only partial orderings. 
Most of the IQA methods are trained on costly perceptual image and annotation pairs,
whereas our approach can be adapted to any image dataset without such labels, and can be applied to any semantic content.

\subsection{Restoration-assisted perceptual IQA}
Several image restoration based approaches have been proposed to generate pseudo-reference images. Most of them rely on GAN-based~\cite{ledig2017photo,baker2002limits,ma2020structure,pan2021exploiting,umer2020deep,wang2018esrgan} or diffusion based methods~\cite{kawar2022denoising,saharia2022image,wang2024exploiting,yang2024pixel} to reconstruct perceptually plausible images. This idea was to adapted to IQA, where restoration networks are used to generate pseudo-references for quality prediction. The method proposed in RAN4IQA~\cite{ren2018ran4iqa} restores distorted images, extracts features from both the distorted and restored images, and evaluates perceptual quality by comparing them. Lin et al. proposed Hallucinated-IQA~\cite{Lin_2018_CVPR}, which uses the discrepancy map between distorted and restored images to predict a quality score. Zheng et al.  CKDN~\cite{zheng2021learning}.  employ a coupled knowledge distillation framework to integrate restoration-related features into IQA. In this way, the model can learn both low-level distortion details and high-level perceptual information. Pan et al. proposed VCRNet~\cite{pan2022vcrnet}, which combines visual compensation and restoration to handle diverse distortions. This design improves the generalization ability of the IQA model. More recently, Chen et al. developed MT-IRN~\cite{chen2023no}. The model jointly learns three tasks: image restoration, distortion type identification, and quality score prediction. It also employs SSIM maps as extra supervision, which guides the model to focus more on structural similarity during training. These restoration-based approaches all aim to predict perceptual quality scores such as MOS. In contrast,
our method uses the processed image and its ISO metadata to estimate a
reference image, from which full-reference metrics such as PSNR, SSIM, and
LPIPS are computed.

\subsection{Camera pipeline}  
Modern digital cameras rely on image signal processing (ISP) pipelines that convert noisy raw sensor measurements into sRGB images. These pipelines include modules for demosaicing~\cite{gharbi2016deep, hasinoff2016burst, wronski2019handheld}, denoising~\cite{zhang2021rethinking, zhang2023practical}, tone mapping~\cite{mertens2007exposure, paris2015local}, and other steps that have been gradually improved over the years. Although these modules enhance perceptual quality, each stage can also introduce distortions~\cite{Renaudin2017eval-multi-image}, which ultimately affect the quality of the final image. In this work, we focus on blindly evaluating such distortions.

\section{Method}

\label{sec:method}

In this section, we describe the proposed framework for evaluating camera pipeline without the need of reference images, as illustrated in Fig.~\ref{fig:diagram}. We begin by modeling the raw image \( x \), which is
subjected to an ISP pipeline \( F \) that transforms it into
an sRGB output \( y = F(x) \). The primary goal of our approach is 
to quantify the artifacts of the pipeline.

The raw image \( x \) is noisy, and it can be modeled as 
a noisy version of an ideal raw measurement \( x_0 \), 
as follows:
\begin{equation}\label{eq:forward_model}
    x = s(x_0 + \varepsilon),
\end{equation}
where \( s \) clips the pixel values between 0 and 1, 
and \( \varepsilon \) represents the noise, typically 
modeled as a Poissonian-Gaussian mixture. The noise 
model depends on the camera sensor and exposure 
conditions, and, importantly, the ideal \( x_0 \) is 
not accessible in practice.
For noise modeling, following
\cite{brooks2019unprocessing,zhang2021rethinking}, we use a
heteroscedastic Gaussian approximation of Poisson--Gaussian noise
with signal-dependent and signal-independent components calibrated
from SIDD noise profiles~\cite{abdelhamed2018dataset}. These
components follow a log-linear relationship, which allows us to use
the ISO value as a proxy for total noise severity.
To evaluate the quality of a processed image \( y \), we would
compare it to a reference image \(\ygt = F(x_0)\), obtained by applying
the camera pipeline \(F\) to the clean raw \(x_0\).
Structural criteria such as PSNR, SSIM, and LPIPS
are then used as $C(y, \ygt)$ to detect subtle differences between
images. While effective, these metrics require perfectly aligned
targets \(\ygt\), which makes it difficult to replicate the exact
acquisition conditions outside controlled laboratory settings. 
A more practical alternative to reference-based evaluation is to use blind 
(or no-reference) methods, which eliminate the need for \( \ygt \).
Instead, we learn a function \( C_\theta(y) \) with parameters $\theta$,
which approximates \( C(y, \ygt) \).
One such strategy, inspired by recent blind IQA
techniques~\cite{ke2021musiq, yang2022maniqa}, is to implement \( C_\theta \)
with a neural network to predict the score \( C(y, \ygt) \) directly from \( y \),
which we refer to as the {\em regression} approach. We show its practical
limits in Section~\ref{sec:experiments}, thus motivating our proposed approach.
In our framework, the network takes as input the processed image \(y\) together
with its ISO value, which is always available in the metadata and serves as a
proxy for the noise level.

\subsection{Camera Pipeline Setting}

\label{sec:pipeline}

Following common practices in computational photography~\cite{brooks2019unprocessing},
we consider a pipeline $F$ that includes white balance, black-and-white level clipping, raw denoising, demosaicing, color conversion, gamma/tone mapping and JPEG compression.
As in Brooks et al.~\cite{brooks2019unprocessing}, the gamma/tone mapping
stage is implemented with a smooth-step curve. For training our method, we use a fixed baseline pipeline where denoising and demosaicing are implemented using the SCUNet~\cite{zhang2023practical}
and Demosaicnet~\cite{gharbi2016deep} neural networks, respectively.
Color conversion is applied with a $3\times3$ matrix, and the white balance,
black and white levels, and color correction matrix for each camera are
obtained from EXIF metadata. In this baseline configuration, tone mapping and
color correction are invertible and do not affect geometric details, so
denoising and demosaicing remain the primary source of distortions. To
further increase realism, we additionally include JPEG compression at
quality factor $Q=90$, which introduces lossy artifacts and leads to a
performance drop reported in Section~\ref{sec:experiments}.

\subsection{Proxy-based Metric Estimation}

Instead of directly regressing metric scores as in the
\emph{regression} approach, we exploit the fact that full-reference
metrics compare two images: the processed image \(y\) and its reference
\(\ygt = F(x_0)\). In practice, pipeline distortions are commonly
quantified by comparing \(y\) with \(\ygt\), which is not available at
inference time. Our idea is to predict a proxy reference image and use
it to approximate \(C(y,\ygt)\). We define our \emph{learnable criterion} \(C_\theta\) as
\begin{equation}
C_\theta(y)
=
C\bigl(y,\net(y,\text{ISO})\bigr).
\end{equation}
When multiple metrics are considered, we use $C^m$ and
$C_\theta^m$ to denote metric $m$ and its proxy-based counterpart,
respectively.
The parametric function \(\net\) is a neural network with weights
\(\theta\) that takes the processed image \(y\) together with the ISO
value, which serves as a proxy for the noise level and sensor
conditions. It returns a proxy reference
\(\yp = \net(y,\text{ISO})\) such that
\(C_\theta(y) \approx C(y,\ygt)\).
In what follows, we omit the ISO dependency for notational clarity.
We call this second strategy the \emph{restoration} approach, aiming
to solve an inverse problem that maps \(y\) to a higher-quality proxy
reference \(\yp\). We implement \(\net\) with a SwinIR
backbone~\cite{liang2021swinir}. To condition on ISO, we replace
LayerNorm in each Transformer block with Adaptive LayerNorm
(AdaLN)~\cite{peebles2023scalable}, where the scale and shift
parameters are predicted from an ISO embedding. This mechanism allows
a single base model to maintain estimation accuracy across a wide
range of noise severities, from low-gain to high-gain scenarios, which
is critical for analyzing diverse ISP configurations.

\begin{figure*}[t]
    \centering
      \includegraphics[width=0.995\linewidth]{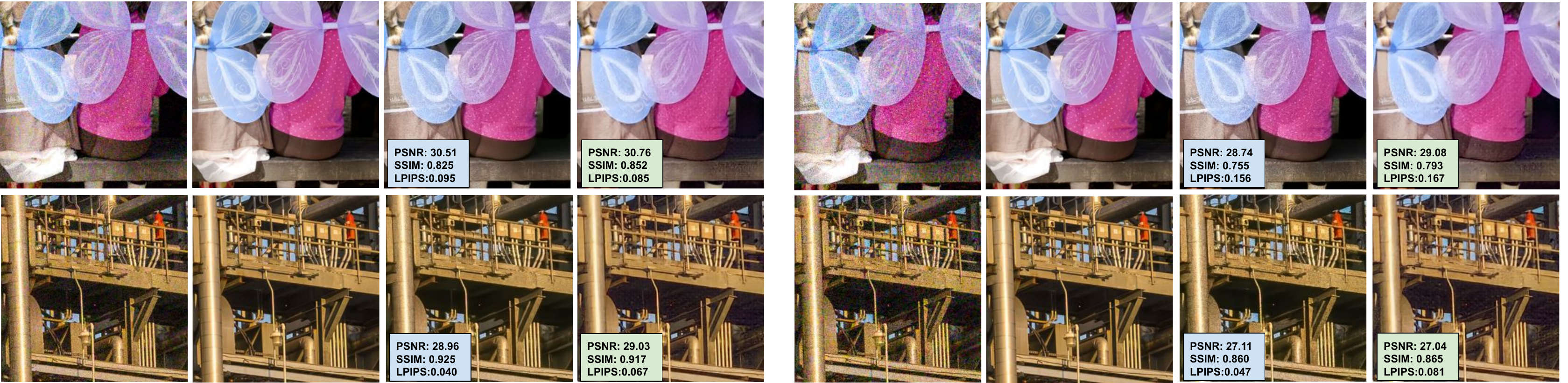}

\vspace{-0.2em}
\centering
\begin{minipage}[t]{0.115\textwidth}
  \centering\small (a) Noisy $x$ \\ {\scriptsize (ISO1600)}

\end{minipage}\hfill
\begin{minipage}[t]{0.115\textwidth}
  \centering\small (b) Reference $\ygt$
\end{minipage}\hfill
\begin{minipage}[t]{0.115\textwidth}
  \centering\small (c) Processed $y$
\end{minipage}\hfill
\begin{minipage}[t]{0.115\textwidth}
  \centering\small (d) Proxy GT
\end{minipage}\hfill
\begin{minipage}[t]{0.115\textwidth}
  \centering\small (e) Noisy $x$ \\ {\scriptsize (ISO3200)}

\end{minipage}\hfill
\begin{minipage}[t]{0.115\textwidth}
  \centering\small (f) Reference $\ygt$
\end{minipage}\hfill
\begin{minipage}[t]{0.115\textwidth}
  \centering\small (g) Processed $y$
\end{minipage}\hfill
\begin{minipage}[t]{0.115\textwidth}
  \centering\small (h) Proxy GT
\end{minipage}
\caption{\textbf{Visual examples of estimated references $y_\theta$.}
Two synthetic patches at ISO1600 and ISO3200 are shown:
noisy inputs $x$, processed outputs $y$, and the corresponding
proxy references predicted by the fine-tuned model $\neta$.
Blue boxes indicate ground-truth metric values,
and green boxes the predicted ones.
Although the estimated references differ from $\ygt$ in color and sharpness, the proxy references yield comparable metric scores.}
\label{fig:visual_example}
\end{figure*}

\subsection{Learning Metric-Specific Reference Images}

A pixel–perfect reconstruction of the GT image $\ygt$
is an ill-posed problem and not required for our task.
Instead, the network only needs to produce an image $\yp$
whose attributes are indistinguishable from those of $\ygt$
under the metrics of interest.
In our case, we consider multiple quality measures, including PSNR,
SSIM, and LPIPS, and aim for $\yp$ to behave similarly to $\ygt$
under these measures.
Each quality metric examines only a specific attribute from the image pair $(y, \ygt)$: PSNR focuses on exact pixel values, SSIM on local structural statistics, and perceptual measures such as LPIPS on deep features.  We directly enforce consistency for PSNR and SSIM, and encourage
perceptual similarity through a feature-based loss.
To do so, we minimize the following distance between the prediction \(\yp\) and the reference \(\ygt\):

\begin{equation}
\ell_\text{data}(y,\ygt)
=
\frac{1}{2}
\bigl\|
\net(y,\text{ISO}) - \ygt
\bigr\|_2^2 .
\end{equation}
\vspace{-0.5em}
Given $N$ training pairs $\left( y^{(i)}, \ygt^{(i)}\right)$,
the base model is trained by minimizing
\begin{equation}
\Loss_{\text{base}}(\theta)
=
\frac{1}{N}\sum_{i=1}^N
\ell_\text{data}\left(y^{(i)}, \ygt^{(i)}\right).
\label{eq:base_loss}
\end{equation} 
Different metrics capture complementary properties
(e.g., PSNR emphasizes pixel-wise fidelity,
SSIM structural similarity,
and LPIPS perceptual similarity).
To align the prediction with specific evaluation metrics,
we introduce a multi-metric consistency loss:

\begin{equation}
\ell_C(y,\ygt)
=
\sum_{m \in \{\text{PSNR},\text{SSIM}\}}
\lambda_m
\left|
C_\theta^m(y) - C^m(y,\ygt)
\right|.
\label{eq:metric_consistency_loss}
\end{equation}
To encourage perceptually plausible outputs,
we further incorporate a perceptual feature loss~\cite{ledig2017photo}:
\begin{equation}
\ell_\text{perc}(\yp,\ygt)
=
\left\|
\phi(\yp) - \phi(\ygt)
\right\|_2^2 ,
\end{equation}
where $\phi$ denotes a pretrained feature extractor. 

The final objective is

\begin{equation}
\begin{aligned}
\Loss(\theta)
=
\frac{1}{N}
\sum_{i=1}^{N}
\Big[
&\ell_{\text{data}}
\left(y^{(i)}, \ygt^{(i)}\right)
+
\ell_C
\left(y^{(i)}, \ygt^{(i)}\right)
\\
&+
\lambda_p
\ell_{\text{perc}}
\left(\yp^{(i)}, \ygt^{(i)}\right)
\Big].
\end{aligned}
\label{eq:overall_loss}
\end{equation}

where $\lambda_{\text{PSNR}}$ and $\lambda_{\text{SSIM}}$
weight the corresponding metric-consistency terms, and
$\lambda_p$ controls the contribution of the perceptual loss.
Qualitative examples of estimated references under this training
objective are shown in Figure~\ref{fig:visual_example}, where noisy
inputs at different ISO levels, their processed outputs and the
proxy references illustrate that multiple visually distinct estimates
can still yield consistent metric scores.

\subsection{Fast Adaptation of Predicting Reference Images}

A practical challenge in evaluating different ISP pipelines is that a single model trained on one pipeline may not generalize well to another, while training separate models from scratch for each pipeline is inefficient. To address this, we follow~\cite{zhang2025augmenting} and use the low-rank adaptation
(LoRA) strategy~\cite{hu2022lora} for fast pipeline-specific adaptation
of \( \net \). LoRA learns a small set of adaptation parameters
\( \omega \) while keeping the base parameters \( \theta \) fixed, and
the adapted model is denoted as \( \neta \).
Our approach consists of two steps: first, we train \( \net \) once on
synthetic data with a pipeline $F$ to obtain general base weights~\( \theta \). Then, for
each target pipeline (e.g., a commercial software ISP such as Adobe
Lightroom, or a pipeline variant with a different denoiser or
demosaicing method), we fine-tune only the LoRA parameters by optimizing
Eq.~\eqref{eq:overall_loss}, while keeping \( \theta \) fixed.
LoRA adapts the model to new
pipelines within a few hours, achieving results comparable to full
fine-tuning while being an order of magnitude faster. This enables
efficient evaluation of multiple ISP pipelines under a fixed noise profile. 
We further show that LoRA fine-tuning also enables fast adaptation when replacing individual components, such as changing demosaicing, denoising, JPEG compression, and PSF-based optical degradation. See Section~\ref{sec:appendix_ISP_Generalization_Results} and Tables~\ref{tab:denoise}, \ref{tab:demosaick_cdmcnn}, \ref{tab:jpeg_full}, and \ref{tab:psf_metric_estimation} in the Appendix.

\section{Experiments}
\label{sec:experiments}
In this section, we evaluate our approach in two settings: (i) controlled synthetic pipelines and (ii) real raw image data processed with Adobe Lightroom. On synthetic evaluation, PSNR, SSIM, and LPIPS are all predicted reliably. However, when applying the synthetic-trained model to Adobe-processed images, PSNR remains stable but SSIM and LPIPS degrade under tone mapping and JPEG. Fine-tuning on the Adobe pipeline restores strong performance across all three metrics.
All of the experiments are run on a single NVIDIA V100 GPU card with
32GB of memory. We implement our code in PyTorch~\cite{paszke2019pytorch}.

\subsection{Evaluation on Synthetic Images}

\paragraph*{Training details.}
We generate synthetic training and validation data using about 8000 clean training images from the DIV2K~\cite{agustsson2017ntire}, Flickr2K~\cite{lim2017enhanced}, and Waterloo~\cite{ma2016waterloo} datasets. From these images, we select $256 \times 256$ non-overlapping crops. To focus on informative texture samples,
we discard smooth patches, {\em e.g.} from out-of-focus
or dynamic regions, following the approach in IQA works
such as NIQE~\cite{mittal2012making}.
We compute a  sharpness coefficient as in NIQE and retain
patches above a sharpness threshold of 0.6,
resulting in 118,135 training patches and 26,487
validation patches. For each patch derived from a
post-processed sRGB image, we reverse the unknown
camera pipeline using the unprocessing strategy from
\cite{brooks2019unprocessing}, which provides the noise-free raw image \( x_0 \),
from which we derive the target processed image \( \ygt \).
We then apply the forward model from
Eq.~\eqref{eq:forward_model}, introducing a noise vector
\( \varepsilon \) sampled from a Poissonian-Gaussian
mixture to produce the noisy raw image \( x \).
The heteroskedastic noise \( \varepsilon \) is modeled as Gaussian noise~\cite{foi2008practical}. 
Its variance depends on the pixel intensity and follows the noise curve \( (\alpha, \beta) \), 
scaled across 8 ISO levels (100, 200, 400, 800, 1600, 3200, 6400, 12800) as calibrated in~\cite{zhang2021rethinking}.
For each
patch, we randomly sample \( (\alpha, \beta) \) by
jittering the calibrated noise curve of a Google Pixel
camera, adapting~\cite{brooks2019unprocessing}.
Lastly, we apply the pipeline \( F \) to the noisy raw image \( x \)
to generate the input image \( y \).
Our synthetic evaluation benchmark consists of 4,162
patches of size $256 \times 256$, taken from 100 test
images in the DIV2K dataset with the same protocol.
The noise in this set also spans ISO values from 100 to 12800.

We train the base model $\net$ from scratch for 90 epochs 
with a batch size of 48, using the AdamW~\cite{loshchilov2017adamw} optimizer 
and a learning rate of \(2 \times 10^{-4}\). 
Random vertical and horizontal flips are applied for data augmentation. 
The base model is trained with the reconstruction loss in 
Eq.~\eqref{eq:base_loss}.  In a second stage, we fine-tune the model using the overall objective 
in Eq.~\eqref{eq:overall_loss}. 
For the joint model, we set
$\lambda_{\text{PSNR}}=0.1$,
$\lambda_{\text{SSIM}}=0.5$, and
$\lambda_p=0.01$.
LoRA adapters are inserted into the attention $q,k,v$ projections, 
MLP layers, and Adaptive LayerNorm modules of each SwinIR block, 
while keeping the backbone frozen. 
In our experiments, adapting only the $q,k,v$ projections produced
visible grid-like artifacts in the estimated references, even though the
metric prediction results were similar. Including the MLP and Adaptive
LayerNorm modules removed these artifacts, and we therefore use this
configuration in all experiments.
After grid-searching the LoRA hyperparameters on synthetic validation
data, we set the rank to 8, the scaling factor to 16, and the dropout
rate to 0.05 for all experiments. 
Fine-tuning is performed for 10 epochs with a batch size of 48 
and a reduced learning rate of \(10^{-4}\). 
This two-stage schedule provides both stable base training 
and efficient adaptation across pipelines.
Additional ablations on LoRA hyperparameters and layer selection are provided in  Table~\ref{tab:lora_ranks}, and Figure~\ref{fig:lora_layers} in Appendix Section~\ref{sec:appendix_Impact_of_LoRA_Layer_Selection}.

\begin{table*}
  \centering
    \caption{\textbf{Estimation of PSNR, SSIM, and LPIPS on synthetic images.}
    Results are obtained using the simplified pipeline $F_{\text{gamma}}$.
    We report mean PLCC, SRCC, and MAE with $\pm$95\% confidence intervals.
    The best result per metric is shown in bold and the second best underlined.
    Restoration models are trained with Eq.~\eqref{eq:overall_loss},
    while the base model uses Eq.~\eqref{eq:base_loss}.}
  \adjustbox{max width=\linewidth}{
  \begin{tabular}{l|ccc|ccc|ccc}
    \toprule
    \multirow{2}{*}{Method} & \multicolumn{3}{c|}{PSNR} & \multicolumn{3}{c|}{SSIM} & \multicolumn{3}{c}{LPIPS} \\
    \cmidrule(l){2-10}
    & PLCC & SRCC & MAE & PLCC & SRCC & MAE & PLCC & SRCC & MAE \\
    \midrule
    Regression 
    & $0.899 \pm 0.002$ & $0.903 \pm 0.002$ & $0.917 \pm 0.009$
    & $0.919 \pm 0.003$ & $0.877 \pm 0.004$ & $0.014 \pm 0.000$
    & $0.587 \pm 0.008$ & $0.614 \pm 0.007$ & $0.040 \pm 0.000$ \\
    \midrule
    Restoration (base)
    & $0.865 \pm 0.003$ & $0.858 \pm 0.004$ & $3.922 \pm 0.019$
    & $0.813 \pm 0.007$ & $0.678 \pm 0.008$ & $0.057 \pm 0.000$
    & $0.682 \pm 0.009$ & $0.697 \pm 0.006$ & $0.075 \pm 0.000$ \\
    Restoration (base) + Perceptual + Adversarial loss
    & $0.892 \pm 0.002$ & $0.898 \pm 0.002$ & $1.733 \pm 0.013$
    & $0.849 \pm 0.006$ & $0.724 \pm 0.007$ & $0.039 \pm 0.000$
    & $0.846 \pm 0.004$ & $0.837 \pm 0.004$ & $0.044 \pm 0.000$ \\
    \midrule
    Restoration ($\lambda_{\text{PSNR}}=0.1$)
    & $\bm{0.969} \pm 0.001$ & $\bm{0.976} \pm 0.001$ & $\bm{0.465} \pm 0.006$
    & $0.906 \pm 0.005$ & $0.875 \pm 0.004$ & $0.025 \pm 0.000$
    & $0.907 \pm 0.004$ & $0.875 \pm 0.004$ & $0.025 \pm 0.000$ \\
    Restoration ($\lambda_{\text{SSIM}}=0.5$) + Perceptual loss
    & $0.960 \pm 0.001$ & $0.964 \pm 0.001$ & $1.219 \pm 0.008$ 
    & $\bm{0.945} \pm 0.003$ & $\bm{0.932} \pm 0.003$ & $\bm{0.009} \pm 0.000$ 
    & $\bm{0.944} \pm 0.003$ & $\bm{0.931} \pm 0.003$ & $\bm{0.009} \pm 0.000$ \\

    Restoration ($\lambda_{\text{LPIPS}}=0.5$)
    & $0.867 \pm 0.003$ & $0.861 \pm 0.004$ & $4.225 \pm 0.022$ 
    & $0.819 \pm 0.007$ & $0.685 \pm 0.008$ & $0.057 \pm 0.000$
    & $0.689 \pm 0.010$ & $0.699 \pm 0.007$ & $0.073 \pm 0.000$ \\

    Restoration ($\lambda_{\text{PSNR}}=0.1$, $\lambda_{\text{SSIM}}=0.5$) + Perceptual loss
    &$\underline{0.966} \pm 0.001$ & $\underline{0.975}
    \pm 0.001$ & $\underline{0.486} \pm 0.006$
    &$\underline{0.943} \pm 0.003$ & $\underline{0.928} \pm 0.003$ & $\bm{0.009} \pm 0.000$
    & $\bm{0.944} \pm 0.003$ & $\underline{0.927} \pm 0.003$ & $\bm{0.009} \pm 0.000$ \\
    
    \bottomrule
  \end{tabular}
  }
\label{tab:restoration_loss_comparison}
\end{table*}

\paragraph*{Estimation of PSNR/SSIM/LPIPS.}

Following common practice in the raw denoising
literature~\cite{brooks2019unprocessing,chen2018learning}, we first report
results with a simplified pipeline $F_{\text{gamma}}$ that ends with gamma
correction (Table~\ref{tab:restoration_loss_comparison}). This configuration
is more linear than the full pipeline and easier to interpret, allowing us
to isolate the effect of different training objectives on metric prediction.
We then extend the evaluation to the full pipeline, which additionally
includes global tone mapping and JPEG compression
(Table~\ref{tab:conditioning_comparison_full_pipeline}). This more realistic
setting better reflects real camera pipelines, but also introduces
non-linear distortions that make reference-free metric estimation
significantly more challenging. Prior work has shown that tone mapping makes
SSIM-based evaluation particularly difficult
\cite{aydin2008extending,yeganeh2012objective,nafchi2014fsitm,
faraji2021full}, and our experiments confirm this trend. JPEG compression
further increases the difficulty by introducing additional lossy artifacts.

Table~\ref{tab:restoration_loss_comparison} compares our
AdaLN-conditioned SwinIR model (75M parameters) with a ViT-AdaLN regression
baseline (154M parameters). Following standard practice in the IQA
community~\cite{ke2021musiq,wang2022exploring,madhusudana2021st,
Su_2020_CVPR,yang2022maniqa,agnolucci2024arniqa}, we evaluate all methods
using Pearson's linear correlation coefficient (PLCC), Spearman's
rank-order correlation coefficient (SRCC), and the mean absolute error
(MAE) between the ground-truth score $C(y,\ygt)$ and the predicted score
$C_\theta(y)$. Each regression entry corresponds to a separate model
trained to predict PSNR, SSIM, or LPIPS directly from $y$.
Direct regression must learn a mapping from the processed image to a single
metric value, although this value depends on the unknown reference image.
Our restoration approach instead predicts a spatially resolved proxy
reference using dense image supervision and then applies the original
full-reference metric between this proxy and $y$. This separates the
estimation of the missing reference from the metric computation and
preserves the pixel, structural, or feature comparisons on which PSNR,
SSIM, and LPIPS are defined.

However, restoration alone does not explain the improvement. The base
restoration model, trained only with the $\ell_\text{data}$ loss
(Eq.~\ref{eq:base_loss}), performs poorly on PSNR and SSIM. The advantage
appears when reconstruction supervision is combined with metric-specific
consistency losses, which guide the proxy reference toward the differences
that are relevant to the target metric.
As is common in image restoration, we also experiment with perceptual and
adversarial losses to encourage photorealistic outputs. The adversarial
loss follows the UNetGAN loss from Real-ESRGAN~\cite{wang2021real}, with
equal weights $(1,1,1)$ for its terms. This setup produces visually better
results, but it reduces the correlation with ground-truth PSNR, SSIM, and
LPIPS scores. In other words, GAN-based training prioritizes perceptual
realism over structural fidelity, making it less suitable for metric
prediction~\cite{blau2018perception}.

When optimizing for SSIM alone, we observed that the restored images often
contained artifacts, suggesting that SSIM by itself is not a sufficient
supervisory signal. To stabilize training, we include an additional
perceptual loss based on VGG features~\cite{ledig2017photo}, which reduces
these artifacts and yields more consistent reconstructions aligned with
the ground-truth SSIM values. In contrast, LPIPS behaves differently.
Training solely with an LPIPS loss
($\lambda_{\text{LPIPS}}=0.5$) leaves SRCC, PLCC, and MAE almost unchanged
from the baseline, indicating that LPIPS provides little useful
regularization on its own. We conjecture that this is because LPIPS
compares abstract deep features that are harder to predict from the
processed image $y$ than the low-level and structural differences captured
by PSNR and SSIM. Adding PSNR or SSIM constraints anchors the optimization
in the structural domain and also reduces LPIPS prediction outliers.

Our metric-weighted variants, such as
$\lambda_{\text{PSNR}}=0.1$ and
$\lambda_{\text{SSIM}}=0.5$ with the VGG loss, achieve performance close to
the best results across all three metrics. In particular, the joint PSNR
and SSIM model performs close to the corresponding metric-specific models
while also maintaining a high correlation with LPIPS. Overall, these
results show that the benefit of the restoration approach comes from
combining dense reconstruction supervision with metric-specific
consistency, rather than from restoration alone. This makes metric
estimation more reliable than direct scalar regression in our setting.

\paragraph*{Impact of noise conditioning.}

We now extend the evaluation to a more practical pipeline that includes
global tone mapping and JPEG compression (\(Q=90\)). Compared to the
simplified gamma-only pipeline, we observe a noticeable performance drop,
especially for SSIM and consequently LPIPS. 
Table~\ref{tab:conditioning_comparison_full_pipeline} compares three
conditioning strategies. In the \emph{no conditioning} setup (first row),
the network receives only the processed image \(y\), which yields the
weakest performance. In the \emph{raw noisy image conditioning} setup
(second row), we additionally provide an auxiliary input obtained by
passing the noisy raw image \(x\) through the same pipeline \(F\) with
denoising disabled. This exposes the model to the noise pattern of the
capture, leading to the best performance, though it requires access to
the raw input. Finally, in the \emph{ISO conditioning} setup (third row),
the model receives \(y\) together with the ISO metadata, serving as a
proxy for noise level. This strategy is slightly less accurate than
conditioning on the raw image, but it is far more practical since ISO
values are readily available in real-world captures. These results highlight that while tone mapping and JPEG compression
make metric prediction harder, especially for SSIM, ISO conditioning offers
a practical balance between accessibility and accuracy.

\begin{table*}
  \centering
    \caption{\textbf{Conditioning comparison under full pipeline $F$ (including tone mapping and JPEG compression with Q=90).}
    Mean PLCC/SRCC/MAE are reported with $\pm$95\% confidence intervals.
    Best is in bold, second best underlined; ISO is our default practical setting. All three methods are trained with perceptual loss.}

  \adjustbox{max width=0.98\linewidth}{
  \begin{tabular}{l|c|ccc|ccc|ccc}
    \toprule
    \multirow{2}{*}{Method} & \multirow{2}{*}{Conditioning} & \multicolumn{3}{c|}{PSNR} & \multicolumn{3}{c|}{SSIM} & \multicolumn{3}{c}{LPIPS} \\
    \cmidrule(l){3-11}
    & & PLCC & SRCC & MAE & PLCC & SRCC & MAE & PLCC & SRCC & MAE \\
    \midrule
    Restoration ($\lambda_{\text{PSNR}}=0.1$, $\lambda_{\text{SSIM}}=0.5$) 
    & NO 
    & $0.923 \pm 0.002$ & $0.940 \pm 0.002$ & $0.591 \pm 0.007$
    & $0.839 \pm 0.008$ & $0.872 \pm 0.005$ & $0.014 \pm 0.000$
    & $0.809 \pm 0.008$ & $0.838 \pm 0.004$ & $0.042 \pm 0.000$ \\
    \midrule
    Restoration ($\lambda_{\text{PSNR}}=0.1$, $\lambda_{\text{SSIM}}=0.5$)
    & Raw noisy image
    & $\bm{0.976} \pm 0.001$ & $\bm{0.987} \pm 0.000$ & $\bm{0.271} \pm 0.005$
    & $\bm{0.933} \pm 0.005$ & $\bm{0.959} \pm 0.003$ & $\bm{0.008} \pm 0.000$
    & $\bm{0.905} \pm 0.005$ & $\bm{0.922} \pm 0.002$ & $\underline{0.025} \pm 0.000$ \\
    \midrule
    Restoration ($\lambda_{\text{PSNR}}=0.1$, $\lambda_{\text{SSIM}}=0.5$)
    & ISO
    & $\underline{0.950} \pm 0.002$ & $\underline{0.967} \pm 0.001$ & $\underline{0.444} \pm 0.007$
    & $\underline{0.869} \pm 0.008$ & $\underline{0.907} \pm 0.005$ & $\underline{0.012} \pm 0.000$
    & $\underline{0.870} \pm 0.008$ & $\underline{0.907} \pm 0.005$ & $\bm{0.012} \pm 0.000$ \\

    \bottomrule
  \end{tabular}
  }
\label{tab:conditioning_comparison_full_pipeline}
\end{table*}

\paragraph*{Comparison to Blind IQA Metrics.}

\begin{figure}[t]
    \centering
    \includegraphics[width=0.9\linewidth]{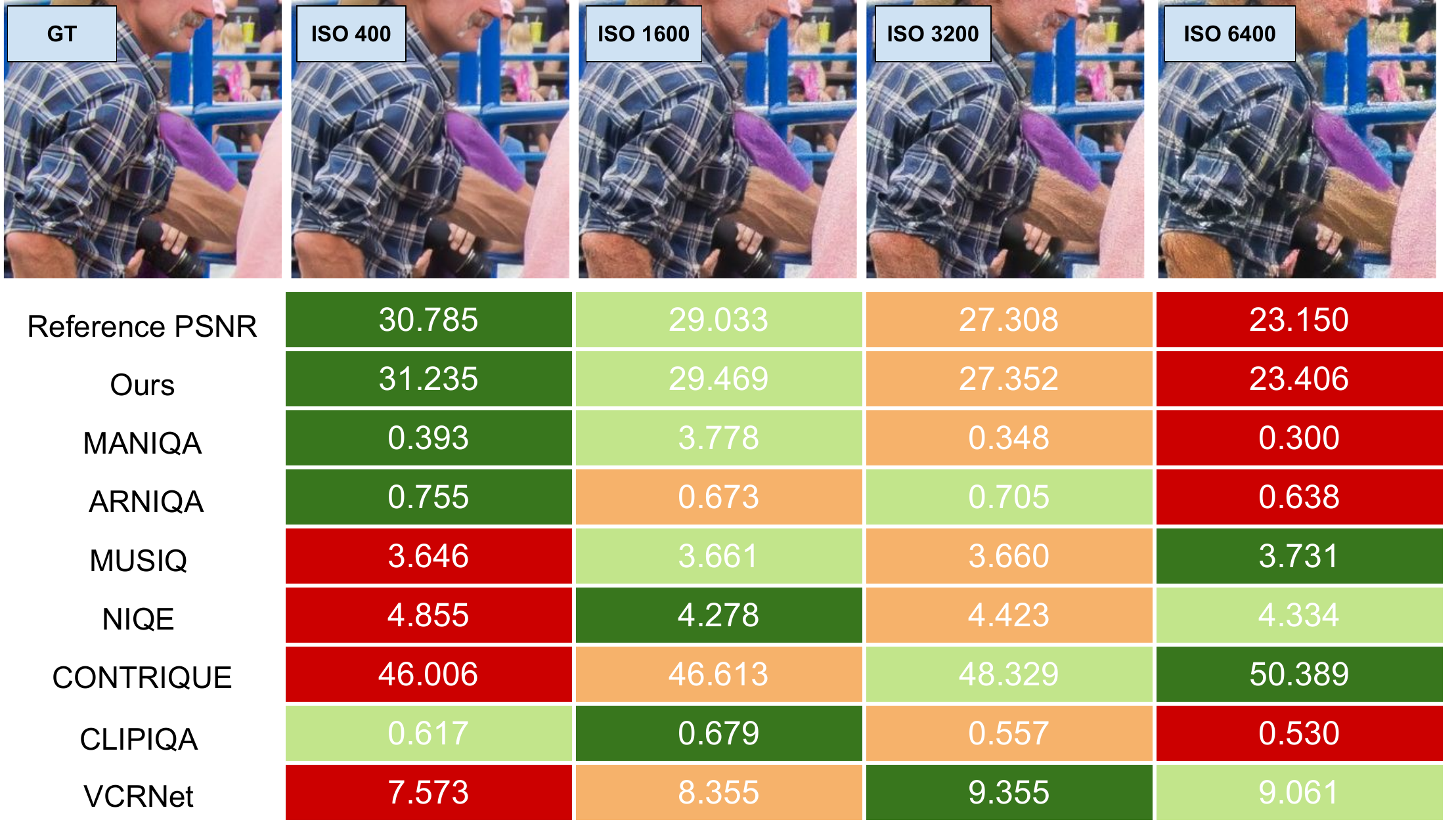}
    \caption{\textbf{Visual example of rankings with blind IQA methods.} Each method computes a quality score for 4 noisy
    versions of a reference image that are degraded with increasing noise levels. Only our method and MANIQA predict the right order, highlighting the insensitivity of blind IQA methods to detect distortions caused by $F$. The reader is invited to zoom in.}
    \label{fig:blind_iqa_ranking}
    \vspace*{-1em}   
\end{figure}

We evaluate our framework's evaluation of ISP artifacts against various blind IQA techniques. Figure~\ref{fig:blind_iqa_ranking} illustrates a representative case where semantic content remains identical while the noise level varies across four ISO settings via $F$. An ideal IQA metric should reliably rank these versions by quality. Although many blind IQA models are specifically trained to estimate denoising artifacts, they often fail to generalize to the realistic, non-linear degradations produced by $F$. In contrast, our restoration-based approach consistently captures the correct quality order (Fig.~\ref{fig:blind_iqa_ranking}), proving that semantic-heavy features are insufficient for evaluating low-level ISP distortions.

Table~\ref{tab:iqa_comparison_task_updated} reports SRCC and PLCC for 10 recent IQA techniques. Our restoration framework consistently achieves the best performance, highlighting its ability to capture low-level distortions that existing blind IQA methods fail to detect. For each clean image~\(\ygt\) in our synthetic evaluation set, we generate
\(20\) raw images \(x\) with random ISO values, process them through the
pipeline \(F\) to obtain \(y\), and compute the ground-truth PSNR, SSIM and
LPIPS rankings of these \(y\) against \(\ygt\). We then compare these true
rankings to the rankings predicted by blind IQA metrics, and our
reference-free PSNR, SSIM and LPIPS estimators applied to the same pools
of candidates. To estimate LPIPS, we use our SSIM estimator as it shows
high correlation with LPIPS. NIQE~\cite{mittal2012making} relies on a statistical model of natural
images that may be too coarse to distinguish small changes introduced by
the ISP. The supervised methods HyperIQA, PAQ2PIQ, DB-CNN, MUSIQ, MANIQA,
and ARNIQA~\cite{Su_2020_CVPR,ying2020patches,zhang2020blind,
ke2021musiq,yang2022maniqa,agnolucci2024arniqa} are trained primarily to
predict subjective quality scores on existing IQA databases. Their learned
quality representations do not necessarily preserve the ordering induced
by individual full-reference metrics on images with identical content and
small differences in ISP processing. Similarly, CONTRIQUE
\cite{madhusudana2021st} and CLIP-IQA~\cite{wang2022exploring} learn
general-purpose perceptual representations, but show limited sensitivity
to the low-level differences considered here. VCRNet~\cite{pan2022vcrnet} is more closely related to our approach, as it
also introduces an image-restoration stage into no-reference IQA. It uses a
non-adversarial restoration network and transfers multi-level restoration
features to a quality-estimation network trained to predict subjective
quality scores. The restoration features therefore provide additional
information for MOS prediction. Since single-image restoration is
ill-posed and is optimized here to support MOS prediction, it does not
necessarily preserve the metric-specific differences measured by PSNR,
SSIM, and LPIPS.

\vspace{-0.2cm}

\begin{table}[t]
\centering
\caption{\textbf{Comparison with IQA metrics.}
Correlation with ground-truth PSNR, SSIM, and LPIPS on the DIV2K
validation set. For each evaluation image, we generate a panel of 20
synthetic noisy/denoised image pairs using the pipeline $F$. We compare
the ranking produced by each IQA method within each panel with the
ground-truth rankings defined by PSNR, SSIM, and LPIPS. The reported
scores are computed over all evaluation images. Our approach achieves the
highest correlation for all three target metrics.}
\adjustbox{max width=0.98\linewidth}{%
\begin{tabular}{lrrrrrr}
\toprule
& \multicolumn{2}{c}{{\small Metric vs PSNR}} 
& \multicolumn{2}{c}{{\small Metric vs SSIM}}
& \multicolumn{2}{c}{{\small Metric vs LPIPS}} \\
 & {\small PLCC} & {\small SRCC} 
 & {\small PLCC} & {\small SRCC} 
 & {\small PLCC} & {\small SRCC} \\
\midrule
{\footnotesize CONTRIQUE~\cite{madhusudana2021st}} 
& -0.563 & -0.546 & -0.533 & -0.535 & -0.461 & -0.416 \\

{\footnotesize CLIPIQA~\cite{wang2022exploring}}       
& 0.147 & 0.276 & 0.117 & 0.275 & 0.084 & 0.246 \\

{\footnotesize DBCNN~\cite{zhang2020blind}}         
& -0.323 & -0.270 & -0.313 & -0.264 & -0.279 & -0.195 \\

{\footnotesize HyperIQA~\cite{Su_2020_CVPR}}      
& 0.054 & 0.097 & 0.052 & 0.100 & 0.040 & 0.105 \\

{\footnotesize MUSIQ~\cite{ke2021musiq}} 
& -0.001 & 0.041 & -0.008 & 0.041 & -0.013 & 0.033 \\

{\footnotesize PAQ2PIQ~\cite{ying2020patches}}       
& 0.147 & 0.276 & 0.117 & 0.275 & 0.084 & 0.246 \\

{\footnotesize NIQE~\cite{mittal2012no}} 
& 0.125 & -0.015 & 0.162 & -0.011 & 0.207 & 0.067 \\

{\footnotesize VCRNet~\cite{pan2022vcrnet}} 
& 0.133 & 0.098 & 0.126 & 0.093 & 0.118 & 0.060 \\

{\footnotesize ARNIQA~\cite{agnolucci2024arniqa}} 
& 0.487 & 0.456 & 0.480 & 0.456 & 0.452 & 0.420 \\

{\footnotesize MANIQA~\cite{yang2022maniqa}}        
& 0.512 & 0.545 & 0.493 & 0.546 & 0.447 & 0.495 \\

{\footnotesize Ours}          
& \textbf{0.993} & \textbf{0.945} 
& \textbf{0.994} & \textbf{0.947} 
& \textbf{0.959} & \textbf{0.833} \\
\bottomrule
\end{tabular}}
\label{tab:iqa_comparison_task_updated}
\end{table}

\paragraph*{Performance of LoRA fine-tuning.}
Table~\ref{tab:lora_comparison} compares training the full network
\(\net\) from scratch and adapting a pre-trained base network via LoRA to obtain
\(\neta\) for PSNR estimation, with \(\lambda_{\mathrm{PSNR}}=0.1\) in
Eq.~\eqref{eq:overall_loss}. Both approaches improve over the base model
trained with $\lambda_{\text{PSNR}}=0$, and achieve comparable PSNR estimation
performance, but LoRA fine-tuning requires only a few hours of adaptation,
versus several days for full training. Thanks to the amortized base training, our approach leads to dramatic reductions in training time when scaling to more metrics or different pipelines.

\begin{table}[t]
    \centering
    \caption{\textbf{LoRA fine-tuning comparison.} 
    A base model is trained from scratch using \eqref{eq:base_loss} 
    ($\lambda_{\text{PSNR}}=0$), a version trained with 
    \eqref{eq:overall_loss} for PSNR estimation ($\lambda_{\text{PSNR}}=0.1$), 
    and a LoRA-adapted base model. The LoRA model is as accurate as the 
    trained one, but requires less training time.}
    \resizebox{0.9\linewidth}{!}{%
    \begin{tabular}{cccc}
        \toprule
        & $\lambda_{\text{PSNR}} = 0$ (Base) & $\lambda_{\text{PSNR}} = 0.1$  & $\lambda_{\text{PSNR}} =0.1$ (LoRA) \\
        & 96 hours & 96 hours &  + 6 hours \\
        \midrule
       PLCC & $0.865\pm0.002$ & $0.955\pm0.001$ & $0.969\pm0.001$ \\
       SRCC & $0.858\pm0.004$ & $0.966\pm0.001$ & $0.976\pm0.001$ \\
       MAE &  $3.922\pm0.019$ & $0.566\pm0.007$ & $0.465\pm0.007$ \\
        \bottomrule
    \end{tabular}}
    \label{tab:lora_comparison}
\end{table}

\subsection{Evaluation on Real Images}
\label{sec:lightroom_eval}

While the previous experiments relied on synthetic clean/noisy pairs, we
now evaluate our method on real data from SIDD
\cite{abdelhamed2018dataset}. Our evaluation set contains approximately
3,000 non-overlapping $256\times256$ patches extracted from 25
noisy/reference 12\,Mpixel raw pairs captured with a Google Pixel (GP).
Additional pairs captured with the LG G4 and Motorola Nexus 6 are used
during LoRA fine-tuning, as described in the following section. The SIDD images differ from the synthetic data used for pretraining due to
the lens PSF, sensor noise characteristics, and scene content. Since our
task involves measuring subtle image differences, this domain gap makes
accurate metric estimation more challenging. The selected images were
captured under neutral illumination, with ISO values ranging from 100 to
10,000. All noisy/reference pairs used in our experiments are processed with the
Adobe Lightroom Adaptive Color pipeline. For evaluation, we extract
approximately 3,000 non-overlapping $256\times256$ patches from the GP
images and retain patches with a sharpness score above 0.2, following the
patch-selection strategy of NIQE~\cite{mittal2012making}. Results obtained
with sharpness thresholds of 0.2, 0.4, and 0.6 are reported in
Appendix~\ref{sec:appendix_synthetic_image_preprocessing}. For zero-shot evaluation, the corresponding SIDD reference images are used
only to compute the ground-truth PSNR, SSIM, and LPIPS values. In the
fine-tuning setting, references from the training split provide supervision
for LoRA adaptation, while references from the held-out GP scenes are used
only for evaluation. In both settings, the model receives only the
processed image and its ISO metadata as input.

\paragraph*{Train/Evaluation setting}
To evaluate direct transfer from the synthetic pipeline to Adobe
Lightroom, we applied the synthetically trained model to the approximately
3,000 GP patches described above, without fine-tuning.
For the Adobe fine-tuned model, we adopted a scene-disjoint protocol
following common practice in denoising works
\cite{zhang2021rethinking}. The GP scenes were divided into three groups:
Group~1 (Scenes 001--003), Group~2 (Scenes 004--006), and Group~3
(Scenes 007--008 and 010). For each fold, one group was reserved for
evaluation, while the remaining scenes were used for LoRA fine-tuning.
To increase the amount of paired fine-tuning data, we also included
noisy/reference pairs from the G4 and N6 subsets. All GP, G4, and N6 pairs
were processed using the same Adobe Lightroom Adaptive Color workflow and
settings. Images corresponding to the held-out scene identities were
excluded from fine-tuning across all devices. The scene split was performed
before patch extraction, ensuring that no patch from a test scene appeared
in the training data. Evaluation was carried out exclusively on patches
from the held-out GP scenes. This setting is reported in
Table~\ref{tab:synthetic_vs_adobe_pipeline} as
\emph{Adobe pipeline (FT)}.

\begin{table*}
  \centering
    \caption{\textbf{Evaluation results with ISO conditioning.}
    We compare the pre-trained the synthetic pipeline $F$ and the model fine-tuned with the Adobe pipeline.
    Evaluation was conducted on Google Phone images from the SIDD dataset, processed with the Adobe Adaptive Color pipeline.
    The Adobe fine-tuned (FT) model was validated with a scene split.}
  \adjustbox{max width=0.98\linewidth}{
  \begin{tabular}{l|ccc|ccc|ccc}
    \toprule
    \multirow{2}{*}{Method} & \multicolumn{3}{c|}{PSNR} & \multicolumn{3}{c|}{SSIM} & \multicolumn{3}{c}{LPIPS} \\
    \cmidrule(l){2-10}
    & PLCC & SRCC & MAE & PLCC & SRCC & MAE & PLCC & SRCC & MAE \\
    \midrule
    Regression
    & $0.750 \pm 0.029$ & $0.751 \pm 0.032$ & $5.823 \pm 0.167$
    &$0.587 \pm 0.057$ & $0.419 \pm 0.052$ & $0.147 \pm 0.008$
    &$0.740 \pm 0.033$ &$0.700 \pm 0.037$ & $0.095 \pm 0.004$ \\
    \midrule
    Synthetic pipeline
    & $0.910 \pm 0.012$ & $0.910 \pm 0.014$ & $4.288 \pm 0.135$
    & $0.656 \pm 0.053$ & $0.459 \pm 0.056$ & $0.124 \pm 0.009$
    & $0.719 \pm 0.038$ & $0.662 \pm 0.037$ & $0.084 \pm 0.005$ \\
    \midrule
    Adobe pipeline (FT)
    & $\bm{0.958} \pm 0.007$ & $\bm{0.942} \pm 0.009$ & $\bm{1.222} \pm 0.087$
    & $\bm{0.957} \pm 0.007$ & $\bm{0.848} \pm 0.027$ & $\bm{0.043} \pm 0.003$
    & $\bm{0.922} \pm 0.011$ & $\bm{0.873} \pm 0.021$ & $\bm{0.050} \pm 0.002$ \\
    \bottomrule
  \end{tabular}
  }
\label{tab:synthetic_vs_adobe_pipeline}
\end{table*}

\paragraph*{Estimation of PSNR, SSIM, and LPIPS.}
Table~\ref{tab:synthetic_vs_adobe_pipeline} reports results for three
settings: (i) regression trained on the synthetic pipeline, (ii) our
restoration model trained on the synthetic pipeline, and (iii) the same
model fine-tuned on Adobe-processed SIDD patches. In general, the restoration approach outperforms regression on PSNR and also improves SSIM over the baseline. We attribute this to the reconstruction objective, which enforces signal-level consistency and provides a more robust inductive bias than direct regression.
The synthetic restoration model still struggles on SSIM before fine-tuning, reflecting a gap between synthetic and real pipelines. 
After fine-tuning on only a few Adobe-processed images with a scene split, the
performance improves substantially: PLCC and SRCC increase across all
metrics, with especially strong gains for SSIM and LPIPS. We hypothesize that synthetic pretraining alone is not sufficient across all metrics due to (i) a domain gap in the raw noise model \((\alpha,\beta)\), which may not accurately match the GP camera and thus produces imperfect synthetic pairs; (ii) residual input--output shifts (e.g., tone mapping or local color adaptation not reproduced by our synthetic ISP); and (iii) limited data diversity due to the small number of training patches and low scene variety. In practice, many outliers occur on text/graphics regions, which are under-represented in our training set.
However, synthetic pretraining provides a good initialization, and LoRA fine-tuning adapts the model to a new pipeline such as Adobe Lightroom with only a few shot images.
This experiment shows that while synthetic pipelines are
useful for pretraining and generalization, real pipelines introduce
nonlinear color and tone variations that cannot be fully captured by
synthetic data. Nonetheless, the ability to bootstrap from a synthetic
pipeline and adapt within hours makes our framework practical for
evaluating commercial ISP pipelines where only limited training data
is available.

\begin{figure*}[!t]
  \centering
  \resizebox{0.8\linewidth}{!}{%
  \begin{subfigure}[t]{0.32\textwidth}
  \includegraphics[width=\linewidth]{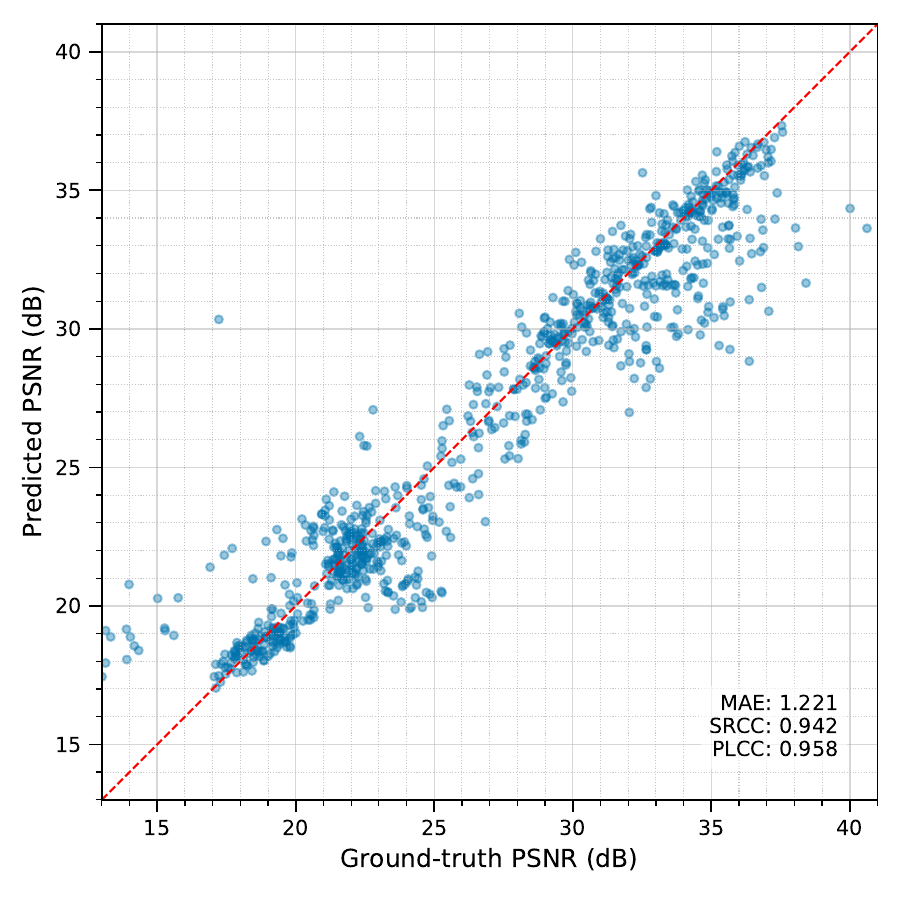}
  \label{fig:triptych-a}
  \end{subfigure}\hfill
  \begin{subfigure}[t]{0.32\textwidth}
  \includegraphics[width=\linewidth]{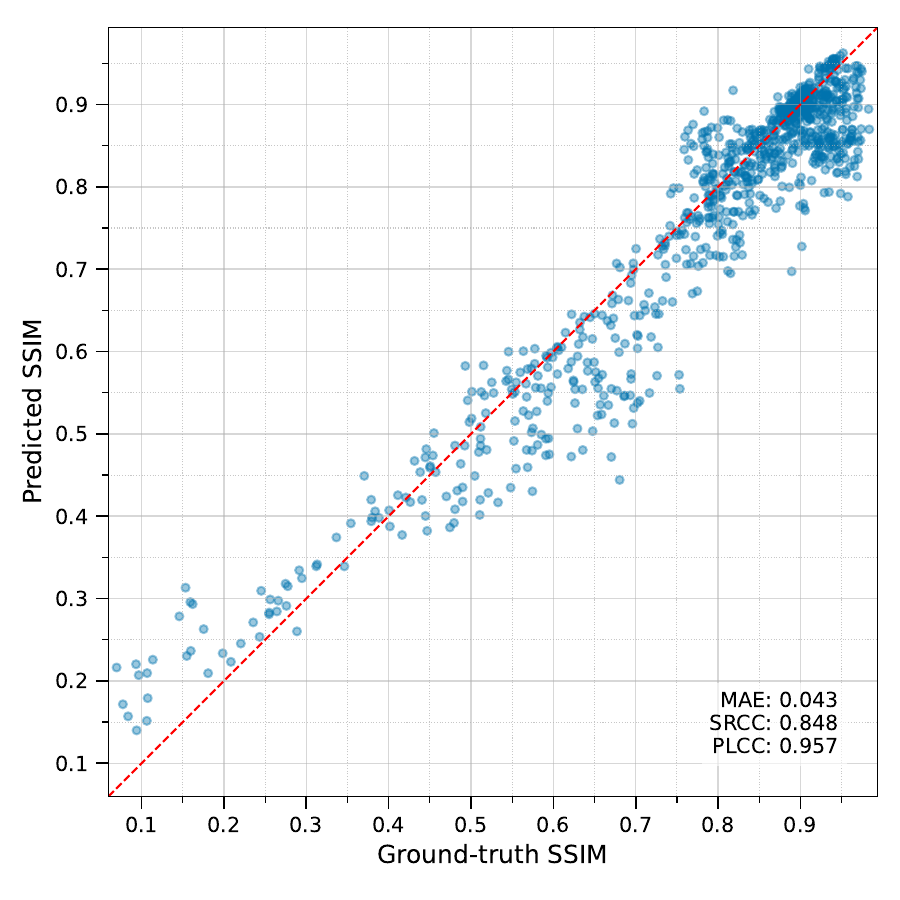}
  \label{fig:triptych-b}
  \end{subfigure}\hfill
  \begin{subfigure}[t]{0.32\textwidth}
  \includegraphics[width=\linewidth]{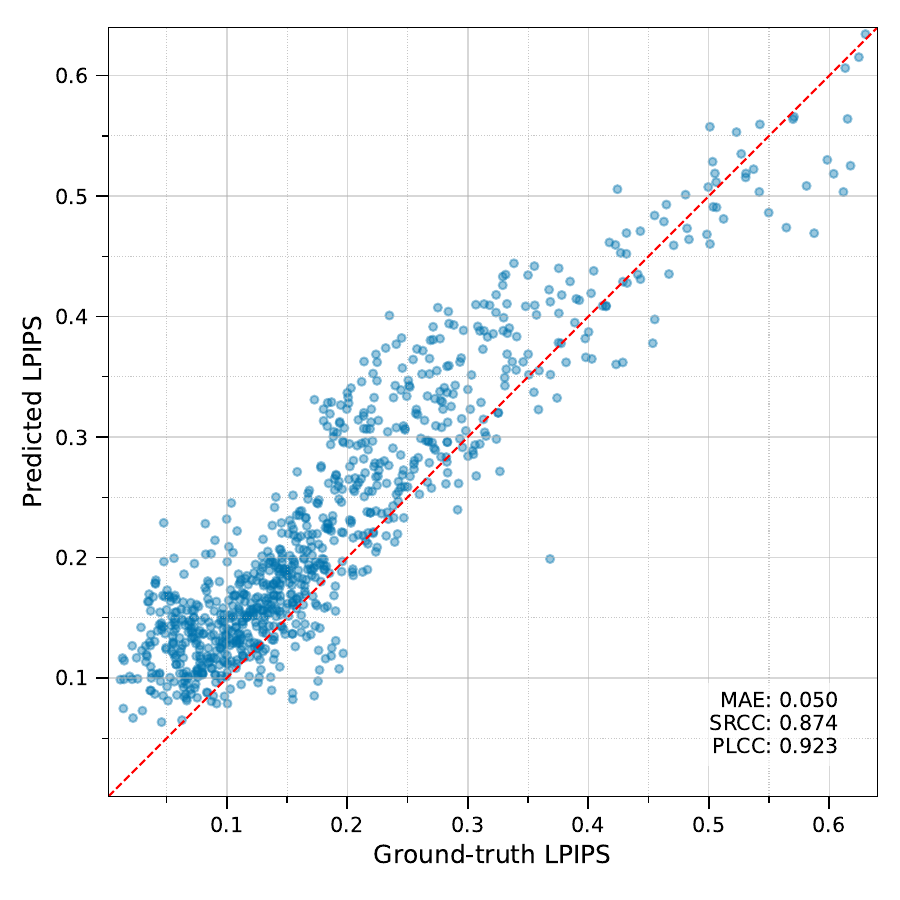}
  \label{fig:triptych-c}
  \end{subfigure}}
  \caption{\textbf{Correlation scores on GP phone patches after Adobe fine-tuning (FT).}
  Scatter plots compare ground-truth values against our predictions for PSNR, SSIM, and LPIPS
  on real pairs of $256\times256$ textured patches taken from SIDD~\cite{abdelhamed2018dataset}.
  The Adobe fine-tuned model attains high Pearson (PLCC) and Spearman (SRCC) correlations across
  all three metrics, indicating effective adaptation of the synthetic-pretrained model to the Adobe pipeline.}
  \label{fig:gp_patch_metrics}
\end{figure*}

\begin{figure*}
\centering
\begin{subfigure}[b]{\linewidth}
\centering
\includegraphics[width=\linewidth]{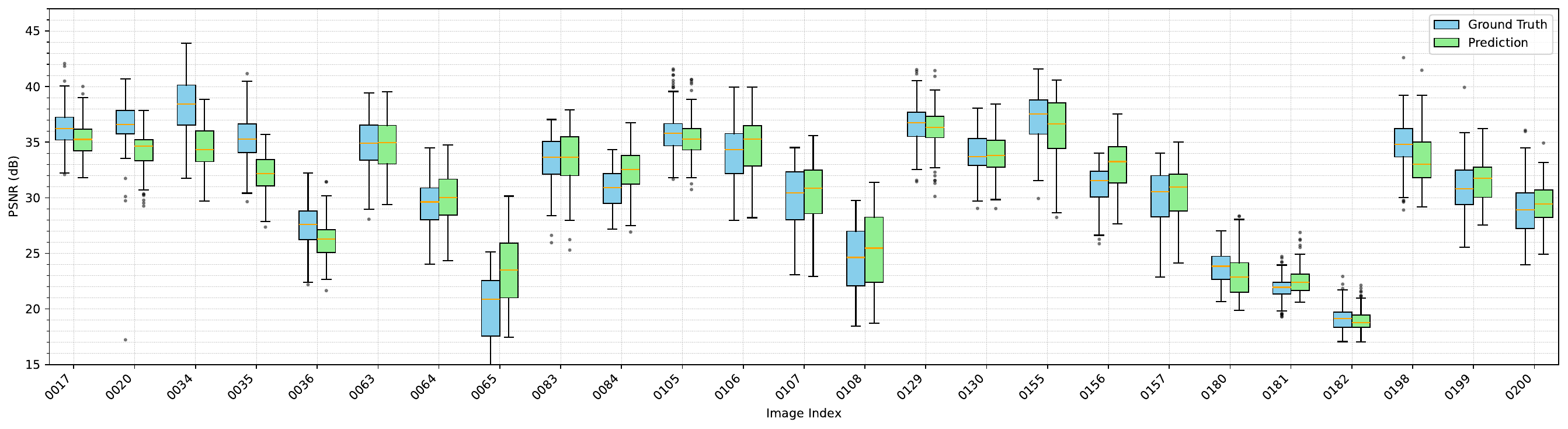}
\caption{Aggregated PSNR results per image show that the predicted medians closely follow the ground-truth medians. In 17 out of 25 cases, the difference between ground-truth and predicted medians remains within 1 dB, while only three images deviate by more than 2 dB. Notably, images \#34, \#35, and \#65 exhibit the largest discrepancies, indicating that the model is generally robust but still struggles on certain challenging scenes.}
\end{subfigure}

\vspace{0.3em}

\begin{subfigure}[b]{\linewidth}
\centering
\includegraphics[width=\linewidth]{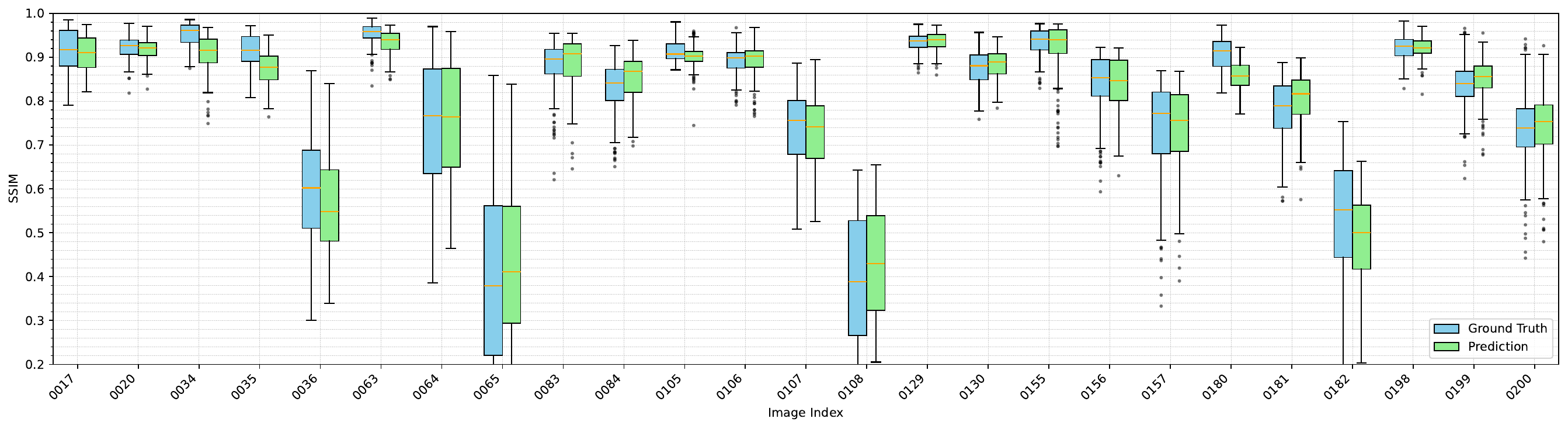}
\caption{Aggregated SSIM results per image show that most predictions are close to the ground-truth medians. However, a subset of images such as \#34, \#36, \#180, and \#182 exhibit deviations greater than 0.05.}
\end{subfigure}

\vspace{0.3em}

\begin{subfigure}[b]{\linewidth}
\centering
\includegraphics[width=\linewidth]{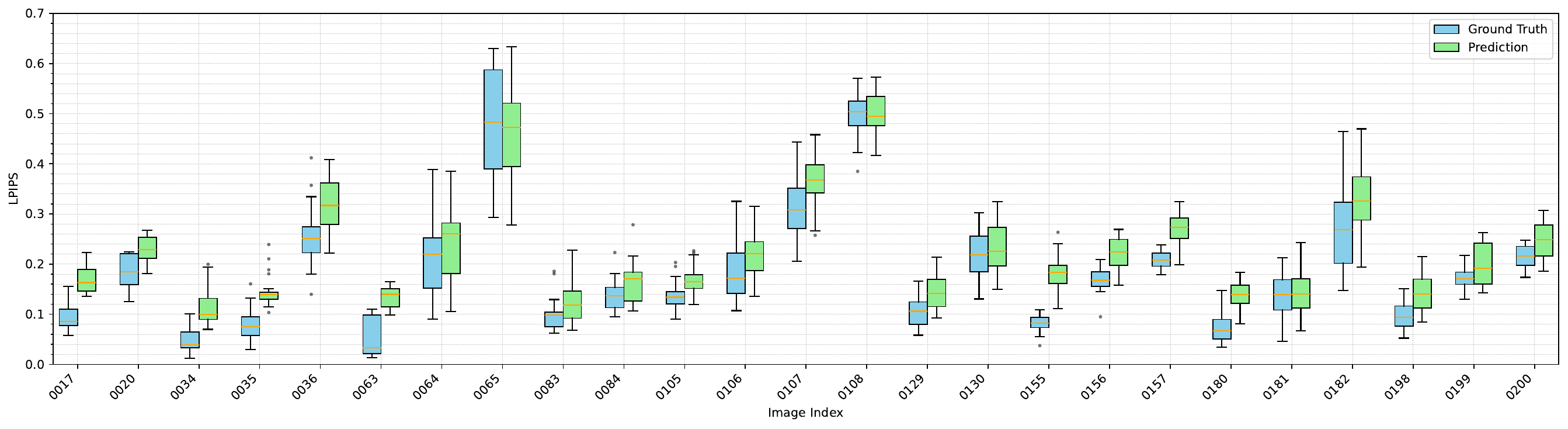}
\caption{Aggregated LPIPS results per image indicate that the predicted medians are generally correlated with the ground truth but often shifted upward, leading to consistent overestimation. Out of 25 images, the majority ($\sim$21) show differences larger than 0.02. Particularly, images \#17, \#34, \#35, and \#36 exhibit larger gaps ($\geq$0.06), highlighting that while the model captures relative trends, it struggles to match the absolute perceptual distances accurately.}
\end{subfigure}

\caption{\textbf{Aggregated quality metric results per image with the Adobe fine-tuned (FT) pipeline.}
We group the patch estimates per image, resulting in 25 boxes, and report medians (orange lines) for PSNR, SSIM, and LPIPS.
These results complement the global correlation analysis in the main paper by highlighting per-image agreement and discrepancies.}
\label{fig:boxplot_adobe}
\end{figure*}

Figure~\ref{fig:gp_patch_metrics} illustrates patch-level correlations
on GP images from the SIDD dataset after Adobe fine-tuning (FT).
The plots confirm that our predicted metrics achieve high PLCC
and SRCC with respect to ground-truth PSNR, SSIM, and LPIPS. Figure~\ref{fig:boxplot_adobe} compares the per-image scores
computed using the ground-truth reference \( \ygt \)
against those obtained with our predicted reference \( \yp \),
using the model fine-tuned on Adobe-processed images (Adobe FT). 
Across PSNR and SSIM, the score distributions exhibit substantial overlap, with closely aligned medians in most cases. For PSNR, 17 out of 25 images show median differences below \(1\,\mathrm{dB}\), and an additional 5 images fall within the \(1\text{--}2\,\mathrm{dB}\) range, leaving only 3 images with deviations larger than \(2\,\mathrm{dB}\). For SSIM, predictions generally follow the ground-truth trends, although a subset of images such as \#34, \#35, \#36, and \#65 shows more noticeable deviations. In contrast, LPIPS results indicate that while relative correlations are well preserved, the predicted medians tend to be systematically shifted upward, reflecting consistent overestimation of perceptual distances.

\section{Conclusion}
\label{ref:conclusion}

We addressed the challenge of evaluating digital camera pipelines by
introducing a reference-free evaluation framework based on processed
sRGB images. Our method takes a processed sRGB image and its ISO value
as input and predicts a proxy reference using a neural network. It then
computes full-reference metrics such as PSNR, SSIM, and LPIPS between
the processed image and the predicted proxy reference.
Our experiments show that existing blind IQA methods do not always
recover the rankings defined by these full-reference metrics when the
image content remains fixed and only the ISP processing changes. In
contrast, our method provides more consistent estimates of the
metric-specific effects of low-level ISP artifacts.
We also introduced a lightweight adaptation strategy based on low-rank
adaptation (LoRA). This allows a pretrained base model to be adapted to
new ISP configurations using a limited amount of data. We showed that
this strategy reduces the domain gap between the synthetic training
pipeline and a realistic RAW-to-sRGB pipeline implemented with Adobe
Lightroom, while retaining the performance of the base model on the
original setting.
In this work, we kept the camera and its noise model fixed in order to
isolate the effects of different denoisers, ISP components, and complete
processing pipelines. Extending the framework to cameras with different
sensor characteristics, as well as to more complex spatially varying
operations such as local contrast enhancement, remains an important
direction for future work.

\section{Acknowledgments}
Work financed by a grant from ANRT N$^\circ$ 2023/0335. This work used GENCI-IDRIS HPC and storage resources under allocations AD011014895R2, AD011012453R4, and AD011017323  on Jean Zay.

\clearpage
\bibliographystyle{eg-alpha-doi} 
\bibliography{references} 

\clearpage
\appendix

\section{IQA dataset distortions}
\label{sec:appendix_IQA_dataset_distortions}
\begin{figure*}[t]
  \centering
  \includegraphics[width=0.98\linewidth]{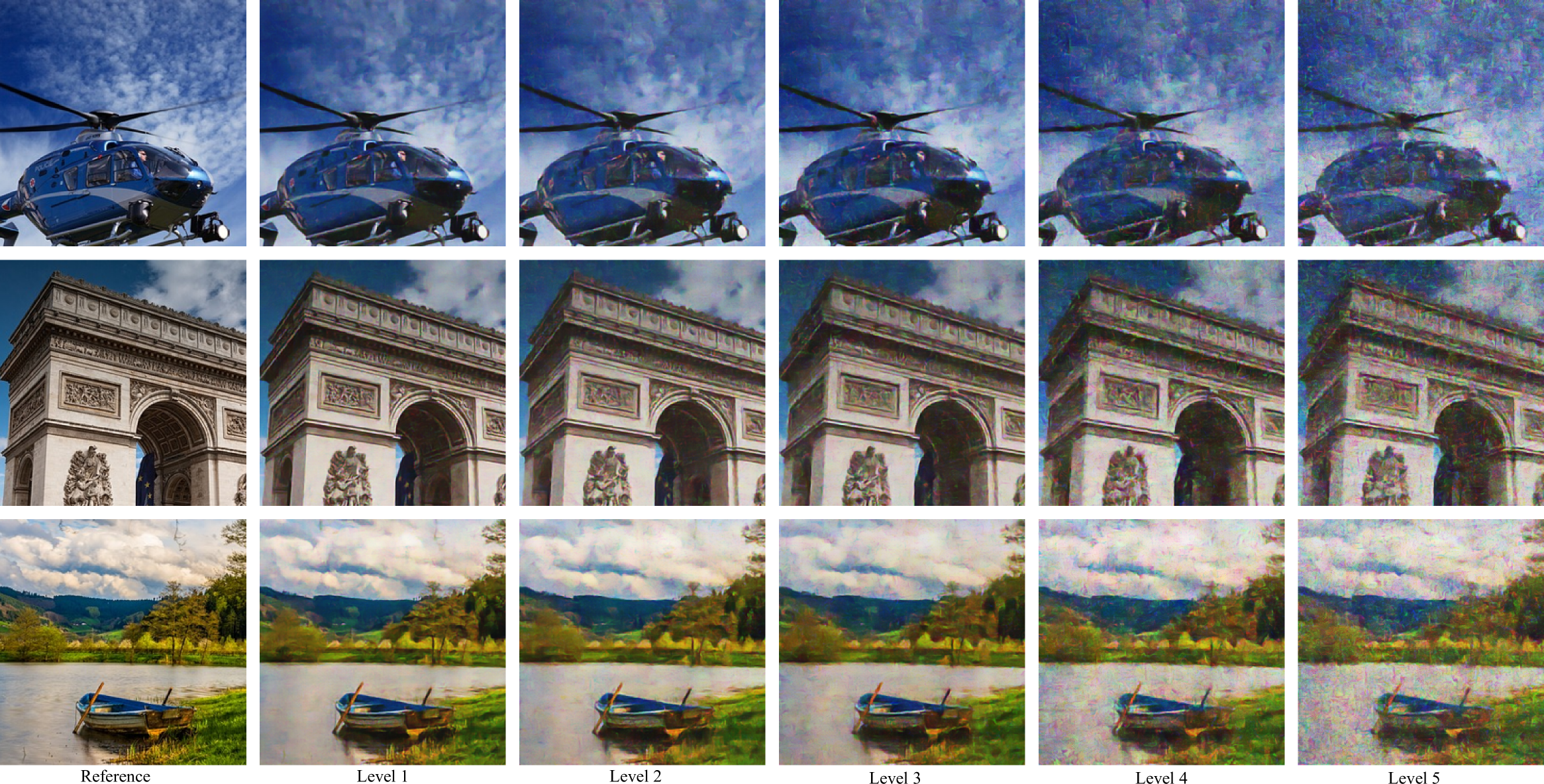}
  \caption{Unrealistic denoising distortions in KADID dataset~\cite{lin2019kadid}.}
  \label{fig:IQA_distortion}
\end{figure*}

There are two types of IQA datasets : authentic distortion datasets and synthetic distortion datasets. Authentic distortion datasets include images affected by realistic conditions such as sensor noise, camera shake, and compression artifacts~\cite{ghadiyaram2015massive, ying2020patches, lin2018koniq, thomee2016yfcc100m, fang2020perceptual}. These datasets capture complex and heterogeneous degradations typically encountered in practical imaging scenarios. However, accurately modeling these authentic distortions is challenging, and the assessment often relies on subjective human evaluations, expressed through Mean Opinion Scores (MOS) or Differential Mean Opinion Scores (DMOS). Synthetic distortion datasets, in contrast, feature artificially induced distortions applied under controlled conditions, including Gaussian noise, JPEG compression, blurring, and other algorithmically-generated degradations~\cite{sheikh2006statistical, ponomarenko2013new, article, lin2019kadid}. Figure~\ref{fig:IQA_distortion} illustrates synthetic denoising distortions from the KADID dataset~\cite{lin2019kadid} at various intensity levels. Nonetheless, such severe synthetic denoising distortions rarely appear in images captured by contemporary cameras.

\section{Training Details}
\label{sec:appendix_Training_Details}
\subsection{Denoiser Training}
For denoiser training, we adopt SCUNet~\cite{zhang2023practical} as our denoiser architecture, originally designed for blind denoising. However, we modify SCUNet to operate in a non-blind configuration by adding a noise-variance derived from noisy input raw image, following Brooks et al.~\cite{brooks2019unprocessing}. Denoising operates on raw data before demosaicing.  We leveraged the calibrated noise profiles from Zhang et al.~\cite{zhang2021rethinking}, derived from the SIDD dataset for Google Pixel and Samsung Galaxy S6 Edge devices. We extrapolated these noise profiles to cover an ISO range of 100 to 12800. Using the same modeling strategy as Brooks et al.~\cite{brooks2019unprocessing}, we established a noise profile for each device, applied a color correction matrix (CCM), and sampled white balance values from the SIDD dataset~\cite{abdelhamed2018dataset} for each device. The denoiser was trained on the DIV2K~\cite{agustsson2017ntire}, Flickr2k~\cite{li2022multi}, and Waterloo Exploration Database~\cite{ma2016waterloo} datasets. We used $512\times512$ patches for training, as larger patch sizes were crucial for maintaining accurate color balance when inverting the gain. During patch processing, $512\times512$ patches were cropped with a stride of 256, including flat patches. We didn't apply Variance Stabilizing Transform (VST) as it can introduce additional errors such as rounding  and clipping. We found that a direct denoising approach without VST provided better image quality in flat regions at very high ISO settings. During the training, noise level is randomly sampled from continuous noise level (ISO100-12800).
The 3x3 color correction matrix is 
proper to each camera model from SIDD dataset~\cite{abdelhamed2018dataset}.

\section{Patch Preprocessing}
\label{sec:appendix_synthetic_image_preprocessing}
We perform patch selection that discards area falling below a sharpness threshold, following the strategy of NIQE~\cite{mittal2012making}. Natural scene statistics-based IQA methods often rely on spatially textured region where distortions
are more reliably detected. By excluding overly flat or out-of-focus patches, we ensure that these statistical models and our restoration metrics remain sensitive to the types of artifacts that meaningfully affect perceived image quality. From processed images with the Adobe lightroom, we selected patches. For this, we used a sharpness threshold $T=0.6$ to discard flat patches, with patch size \(256 \times 256\) with stride=256. We can see the performance is quite robust even we lower down the threshold and allow more patches in Table~\ref{tab:swinir_threshold}.

\begin{table}[h]
    \centering
    \caption{\textbf{Impact of different sharpness thresholds for patch selection on PSNR estimation evaluated using the SIDD Google Phone dataset.~\cite{abdelhamed2018dataset}}  We report mean PLCC, SRCC, and MAE accompanied by $\pm$95\% confidence intervals.}
    \begin{tabular}{cccc}
    \toprule
    Threshold & PLCC & SRCC & MAE \\
    \midrule
    0.2 & $0.946 \pm 0.004$ & $0.930 \pm 0.006$& $1.441 \pm 0.043$\\
    0.4 &$0.955 \pm 0.004$ &$0.941 \pm 0.007$ &$1.316 \pm 0.057$\\
    0.6 &$0.958 \pm 0.007$ & $0.942 \pm 0.009$ &$1.219 \pm 0.083$ \\
    \bottomrule
    \end{tabular}
    \label{tab:swinir_threshold}
\end{table}

\section{Regression vs.\ Restoration Training}

Both models take the processed image $y$ and its ISO metadata as input,
but differ in their output and supervision. The restoration model predicts
a proxy reference $\net(y,\mathrm{ISO})$ and is trained using the objective
in Eq.~\eqref{eq:overall_loss}, which combines reconstruction,
metric-consistency, and, when used, perceptual losses.

The regression baseline instead predicts the target metric directly.
For a metric $C^m$, we denote the regression model by $R_\psi^m$ and
optimize

\begin{equation}
\label{eq:regression_loss}
\Loss_{\mathrm{reg}}^m
=
\frac{1}{N}
\sum_{i=1}^{N}
\left|
R_\psi^m
\left(
y^{(i)},\mathrm{ISO}^{(i)}
\right)
-
C^m
\left(
y^{(i)},\ygt^{(i)}
\right)
\right|.
\end{equation}

Thus, both approaches use the same processed image and ISO information.
The regression model is supervised only by one scalar metric value per
patch, whereas the restoration model additionally receives dense
supervision from the reference image and preserves the original two-image
computation of the metric.

\section{ISP Generalization Results}
\label{sec:appendix_ISP_Generalization_Results}

Although the predictor was trained on a single ISP configuration $F$ (SCUNet denoising $\rightarrow$ DemosaicNet $\rightarrow\gamma\!=\!2.2$, smooth-step curve and JPEG compression Q=90). The trained estimator remain highly reliable even when each stage is replaced at test time. We present the mean PLCC, SRCC, and MAE for PSNR estimation, accompanied by $\pm$95\% confidence intervals in Table~\ref{tab:denoise},~\ref{tab:demosaick_cdmcnn}.

\vspace{0.5cm}
\noindent\textbf{Changing the denoiser.}
Table~\ref{tab:denoise} replaces SCUNet~\cite{zhang2023practical} with DRUNet~\cite{zhang2021plug} at test time. Using different denoiser gives different residuals and artifacts. 
Without any adaptation, correlations remains around 0.86-0.88. However, the distributional shift in noise statistics degrades calibration and ranking, especially at higher ISO. The model finetuning with the DRUNet (DRUNet~FT) boost the performance,  indicating that it can be quickly re-aligned to a new denoiser.

\vspace{0.5cm}
\noindent\textbf{Changing the demosaicing method.}
Changing the demosaicing method also defines a new ISP pipeline, since different algorithms employ different strategies to reconstruct edges, textures, and color details from the Bayer input.
We evaluate this effect by replacing DemosaicNet~\cite{gharbi2016deep} with the CNN-based demosaicing method CDM-CNN~\cite{tan2017color}, and by subsequently fine-tuning the estimator on this new pipeline configuration.
The results in Table~\ref{tab:demosaick_cdmcnn} show that, even without fine-tuning, the predictor remains robust, although the errors increase compared with those obtained using the demosaicing method seen during training.
After fine-tuning, the predictor rapidly adapts and achieves performance close to that obtained on the original pipeline.
These results confirm that the choice of demosaicing method, similarly to the choice of denoising method, directly affects the final image quality, while such pipeline changes can be simulated and accommodated within our framework.

\vspace{0.5cm}
\noindent\textbf{Generalization to JPEG compression.}
Table~\ref{tab:jpeg_full} evaluates distribution shift induced by
JPEG compression by comparing in-distribution performance (Q=90), zero-shot
performance at Q=50 and Q=10, and adaptation with lightweight LoRA fine-tuning
at Q=10. Zero-shot results degrade as compression becomes stronger: average
PLCC drops from $0.952$ (Q=90) to $0.922$ (Q=50) and $0.777$ (Q=10), while MAE
increases from $0.442$ to $2.721$ and $6.364$, respectively. After LoRA
fine-tuning on Q=10, performance recovers strongly (average PLCC $0.983$,
SRCC $0.987$, MAE $0.316$), showing that the predictor is robust to severe
JPEG artifacts once adapted with a small amount of target-domain data.

\vspace{0.5cm}
\noindent\textbf{Generalization to optical degradation (PSF)}
To evaluate robustness to optical degradations, we follow the approach of Eboli et al.~\cite{eboli22fast} and use PSFs estimated from real camera data. These estimated PSFs are applied to our synthetic pipeline to generate optically degraded data under realistic aberrations. The results in Table~\ref{tab:psf_metric_estimation} show that metric prediction remains robust across ISO levels, indicating that the model generalizes well even when synthetic data are augmented with real-camera optical blur.

\vspace{0.5cm}

 \begin{figure*}[t!]
    \centering
    \includegraphics[width=0.99\linewidth]{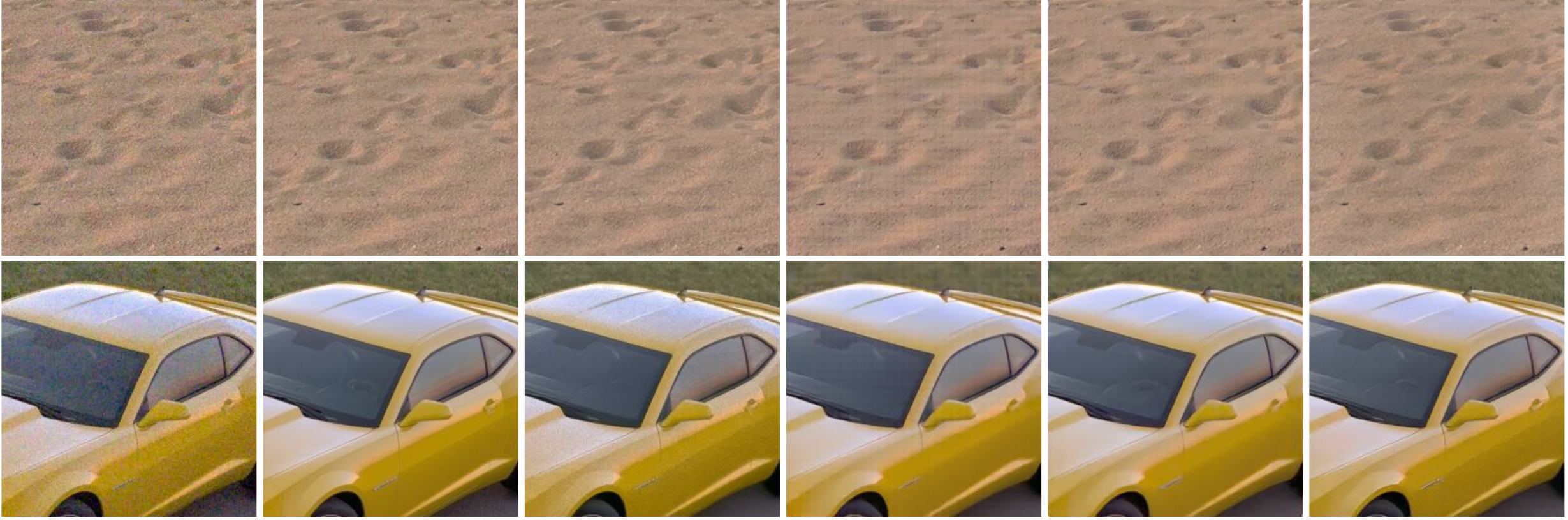}

\vspace{0.3cm}
\begin{minipage}[t]{0.16\textwidth}
  \centering\small (a) Noisy $x$ \\ {\scriptsize (ISO1600)}
\end{minipage}\hfill
\begin{minipage}[t]{0.16\textwidth}
  \centering\small (b) Reference $\ygt$
\end{minipage}\hfill
\begin{minipage}[t]{0.16\textwidth}
  \centering\small (c) Processed $y$
\end{minipage}\hfill
\begin{minipage}[t]{0.16\textwidth}
  \centering\small (d) Attn $q,k,v$
\end{minipage}\hfill
\begin{minipage}[t]{0.16\textwidth}
  \centering\small (e) Attn $q,k,v$, MLP, \\ AdaLN layers
\end{minipage}\hfill
\begin{minipage}[t]{0.16\textwidth}
  \centering\small (f) Attn $q,k,v$, MLP, \\ AdaLN layers + VGG loss
\end{minipage}

\caption{\textbf{Comparison of different LoRA configurations.} 
Six variants are shown: (a) noisy raw $x$, (b) ground-truth reference $\ygt$, (c) processed image $y$, (d) fine-tuning only attention $q,k,v$, (e) attention + MLP + AdaLN layers, and (f) attention + MLP + AdaLN layers with VGG perceptual loss.}
\label{fig:lora_layers}
\end{figure*}

\begin{table*}[p]
  \centering
  \caption{\textbf{Impact of different denoising methods on PSNR estimation evaluated using synthetic raw images.} SCUNet (trained) corresponds to our default ISP pipeline configuration. DRUNet denotes evaluation when only the denoiser is replaced with DRUNet~\cite{zhang2021plug}, without any re-training. DRUNet (FT) indicates the model fine-tuned on synthetic data with DRUNet as the denoiser. We show evaluation results without/with finetuning. We report the mean PLCC, SRCC, and MAE metrics with $\pm95$ confidence intervals.}
  \label{tab:denoise}
  \begin{adjustbox}{max width=\linewidth}
  \begin{tabular}{c|ccc|ccc|ccc}
    \toprule
    \multirow{2}{*}{\textbf{ISO}}
      & \multicolumn{3}{c|}{SCUNet (trained)}
      & \multicolumn{3}{c|}{DRUNet}
      & \multicolumn{3}{c}{DRUNet (FT)} \\
    \cmidrule(lr){2-4}\cmidrule(lr){5-7}\cmidrule(lr){8-10}
      & PLCC & SRCC & MAE & PLCC & SRCC & MAE & PLCC & SRCC & MAE \\
    \midrule
    100  & 0.920$\pm$0.010 & 0.956$\pm$0.003 & 0.590$\pm$0.024
         & 0.872$\pm$0.009 & 0.854$\pm$0.010 & 0.862$\pm$0.021
         & 0.975$\pm$0.002 & 0.974$\pm$0.002 & 0.341$\pm$0.010 \\
    200  & 0.929$\pm$0.010 & 0.960$\pm$0.003 & 0.561$\pm$0.021
         & 0.875$\pm$0.009 & 0.855$\pm$0.009 & 0.943$\pm$0.021
         & 0.975$\pm$0.002 & 0.973$\pm$0.002 & 0.335$\pm$0.009 \\
    400  & 0.946$\pm$0.007 & 0.962$\pm$0.003 & 0.514$\pm$0.017
         & 0.881$\pm$0.010 & 0.861$\pm$0.009 & 1.153$\pm$0.021
         & 0.976$\pm$0.002 & 0.974$\pm$0.002 & 0.313$\pm$0.009 \\
    800  & 0.958$\pm$0.003 & 0.966$\pm$0.002 & 0.463$\pm$0.013
         & 0.890$\pm$0.008 & 0.868$\pm$0.009 & 1.548$\pm$0.022
         & 0.976$\pm$0.002 & 0.973$\pm$0.002 & 0.286$\pm$0.009 \\
    1600 & 0.964$\pm$0.002 & 0.969$\pm$0.002 & 0.373$\pm$0.011
         & 0.900$\pm$0.007 & 0.879$\pm$0.009 & 2.091$\pm$0.020
         & 0.977$\pm$0.002 & 0.974$\pm$0.002 & 0.246$\pm$0.007 \\
    3200 & 0.968$\pm$0.002 & 0.976$\pm$0.002 & 0.304$\pm$0.010
         & 0.905$\pm$0.006 & 0.892$\pm$0.008 & 2.427$\pm$0.019
         & 0.980$\pm$0.002 & 0.980$\pm$0.002 & 0.208$\pm$0.007 \\
    6400 & 0.978$\pm$0.002 & 0.977$\pm$0.002 & 0.285$\pm$0.008
         & 0.847$\pm$0.010 & 0.810$\pm$0.014 & 3.079$\pm$0.023
         & 0.982$\pm$0.002 & 0.983$\pm$0.002 & 0.181$\pm$0.006 \\
    \midrule
    Average
         & 0.952$\pm$0.002 & 0.967$\pm$0.001 & 0.442$\pm$0.006
         & 0.881$\pm$0.003 & 0.860$\pm$0.004 & 1.729$\pm$0.008
         & 0.977$\pm$0.001 & 0.976$\pm$0.001 & 0.273$\pm$0.003 \\
    \bottomrule
  \end{tabular}
  \end{adjustbox}
\end{table*}

\begin{table*}[p]
  \centering
  \caption{\textbf{Impact of different demosaicing methods on PSNR estimation evaluated using synthetic raw images.} Here we compare with the recent CNN based demosaicing methods (CDM-CNN)~\cite{tan2017color}, and shows evaluation results without/with finetuning. We report the mean PLCC, SRCC, and MAE metrics with $\pm95$ confidence intervals.}
  \label{tab:demosaick_cdmcnn}
  \begin{adjustbox}{max width=\linewidth}
  \begin{tabular}{c|ccc|ccc|ccc}
    \toprule
    \multirow{2}{*}{\textbf{ISO}}
      & \multicolumn{3}{c|}{DemosaicNet (trained)}
      & \multicolumn{3}{c|}{Cdmcnn}
      & \multicolumn{3}{c}{Cdmcnn (FT)} \\
    \cmidrule(lr){2-4}\cmidrule(lr){5-7}\cmidrule(lr){8-10}
      & PLCC & SRCC & MAE & PLCC & SRCC & MAE & PLCC & SRCC & MAE \\
    \midrule
    100  & 0.920$\pm$0.010 & 0.956$\pm$0.003 & 0.590$\pm$0.024
         & 0.924$\pm$0.006 & 0.938$\pm$0.004 & 2.249$\pm$0.031
         & 0.917$\pm$0.008 & 0.947$\pm$0.003 & 0.922$\pm$0.024 \\
    200  & 0.929$\pm$0.010 & 0.960$\pm$0.003 & 0.561$\pm$0.021
         & 0.932$\pm$0.005 & 0.937$\pm$0.004 & 2.294$\pm$0.029
         & 0.923$\pm$0.007 & 0.947$\pm$0.003 & 0.907$\pm$0.022 \\
    400  & 0.946$\pm$0.007 & 0.962$\pm$0.003 & 0.514$\pm$0.017
         & 0.941$\pm$0.004 & 0.942$\pm$0.004 & 2.371$\pm$0.025
         & 0.931$\pm$0.006 & 0.951$\pm$0.003 & 0.906$\pm$0.020 \\
    800  & 0.958$\pm$0.003 & 0.966$\pm$0.002 & 0.463$\pm$0.013
         & 0.941$\pm$0.004 & 0.939$\pm$0.005 & 2.479$\pm$0.024
         & 0.941$\pm$0.004 & 0.956$\pm$0.003 & 0.893$\pm$0.018 \\
    1600 & 0.964$\pm$0.002 & 0.969$\pm$0.002 & 0.373$\pm$0.011
         & 0.934$\pm$0.004 & 0.930$\pm$0.005 & 2.449$\pm$0.021
         & 0.951$\pm$0.004 & 0.964$\pm$0.003 & 0.813$\pm$0.016 \\
    3200 & 0.968$\pm$0.002 & 0.976$\pm$0.002 & 0.304$\pm$0.010
         & 0.920$\pm$0.005 & 0.909$\pm$0.007 & 2.235$\pm$0.022
         & 0.954$\pm$0.004 & 0.968$\pm$0.002 & 0.810$\pm$0.016 \\
    6400 & 0.978$\pm$0.002 & 0.977$\pm$0.002 & 0.285$\pm$0.008
         & 0.899$\pm$0.006 & 0.888$\pm$0.007 & 1.826$\pm$0.025
         & 0.966$\pm$0.003 & 0.970$\pm$0.002 & 0.905$\pm$0.016 \\
    \midrule
    Average
         & 0.952$\pm$0.002 & 0.967$\pm$0.001 & 0.442$\pm$0.006
         & 0.927$\pm$0.002 & 0.926$\pm$0.002 & 2.272$\pm$0.010
         & 0.940$\pm$0.002 & 0.958$\pm$0.001 & 0.879$\pm$0.007 \\
    \bottomrule
  \end{tabular}
  \end{adjustbox}
\end{table*}

\begin{table*}[h]
\centering
\caption{\textbf{Impact of JPEG compression on PSNR estimation.}
The base model is trained with JPEG Q=90.
We report in-distribution performance (Q=90), zero-shot evaluation under moderate (Q=50)
and severe (Q=10) compression, and performance after lightweight LoRA fine-tuning on Q=10 data.
All results are reported as mean $\pm$ 95\% confidence intervals.}
\label{tab:jpeg_full}
\begin{adjustbox}{max width=\linewidth}
\begin{tabular}{c|ccc|ccc|ccc|ccc}
\toprule
\multirow{2}{*}{\textbf{ISO}}
& \multicolumn{3}{c|}{Q=90 (In-distribution)}
& \multicolumn{3}{c|}{Q=50 (Zero-shot)}
& \multicolumn{3}{c|}{Q=10 (Zero-shot)}
& \multicolumn{3}{c}{Q=10 (LoRA FT)} \\
\cmidrule(lr){2-4}
\cmidrule(lr){5-7}
\cmidrule(lr){8-10}
\cmidrule(lr){11-13}
& PLCC & SRCC & MAE
& PLCC & SRCC & MAE
& PLCC & SRCC & MAE
& PLCC & SRCC & MAE \\
\midrule
100  & 0.920$\pm$0.010 & 0.956$\pm$0.003 & 0.590$\pm$0.024
     & 0.888$\pm$0.006 & 0.896$\pm$0.006 & 2.970$\pm$0.036
     & 0.729$\pm$0.015 & 0.668$\pm$0.019 & 7.284$\pm$0.052
     & 0.983$\pm$0.001 & 0.988$\pm$0.001 & 0.320$\pm$0.011 \\

200  & 0.929$\pm$0.010 & 0.960$\pm$0.003 & 0.561$\pm$0.021
     & 0.893$\pm$0.006 & 0.900$\pm$0.006 & 2.945$\pm$0.036
     & 0.733$\pm$0.015 & 0.672$\pm$0.019 & 7.217$\pm$0.054
     & 0.983$\pm$0.001 & 0.988$\pm$0.001 & 0.319$\pm$0.010 \\

400  & 0.946$\pm$0.007 & 0.962$\pm$0.003 & 0.514$\pm$0.017
     & 0.901$\pm$0.006 & 0.910$\pm$0.006 & 2.892$\pm$0.035
     & 0.742$\pm$0.014 & 0.681$\pm$0.018 & 7.079$\pm$0.051
     & 0.983$\pm$0.001 & 0.987$\pm$0.001 & 0.324$\pm$0.011 \\

800  & 0.958$\pm$0.003 & 0.966$\pm$0.002 & 0.463$\pm$0.013
     & 0.919$\pm$0.005 & 0.926$\pm$0.005 & 2.783$\pm$0.030
     & 0.760$\pm$0.014 & 0.699$\pm$0.019 & 6.815$\pm$0.048
     & 0.984$\pm$0.001 & 0.988$\pm$0.001 & 0.317$\pm$0.010 \\

1600 & 0.964$\pm$0.002 & 0.969$\pm$0.002 & 0.373$\pm$0.011
     & 0.946$\pm$0.003 & 0.945$\pm$0.004 & 2.698$\pm$0.025
     & 0.792$\pm$0.013 & 0.737$\pm$0.017 & 6.372$\pm$0.045
     & 0.983$\pm$0.001 & 0.987$\pm$0.001 & 0.314$\pm$0.011 \\

3200 & 0.968$\pm$0.002 & 0.976$\pm$0.002 & 0.304$\pm$0.010
     & 0.959$\pm$0.002 & 0.956$\pm$0.003 & 2.530$\pm$0.020
     & 0.830$\pm$0.011 & 0.784$\pm$0.014 & 5.602$\pm$0.040
     & 0.985$\pm$0.001 & 0.988$\pm$0.001 & 0.300$\pm$0.009 \\

6400 & 0.978$\pm$0.002 & 0.977$\pm$0.002 & 0.285$\pm$0.008
     & 0.951$\pm$0.003 & 0.943$\pm$0.004 & 2.229$\pm$0.021
     & 0.853$\pm$0.009 & 0.813$\pm$0.013 & 4.179$\pm$0.036
     & 0.981$\pm$0.001 & 0.983$\pm$0.001 & 0.320$\pm$0.009 \\

\midrule
Average
     & 0.952$\pm$0.002 & 0.967$\pm$0.001 & 0.442$\pm$0.006
     & 0.922$\pm$0.002 & 0.925$\pm$0.002 & 2.721$\pm$0.011
     & 0.777$\pm$0.005 & 0.722$\pm$0.006 & 6.364$\pm$0.018
     & 0.983$\pm$0.001 & 0.987$\pm$0.000 & 0.316$\pm$0.004 \\
\bottomrule
\end{tabular}
\end{adjustbox}
\end{table*}

\begin{table*}[p]
\centering
\caption{\textbf{Evaluation results with PSF.~\cite{eboli22fast}}
PSNR, SSIM, and LPIPS are computed between the degraded image and the restored output.
We report PLCC, SRCC, and MAE w.r.t.\ ground-truth full-reference metrics.}
\label{tab:psf_metric_estimation}
\begin{adjustbox}{max width=\linewidth}
\begin{tabular}{c|ccc|ccc|ccc}
\toprule
\multirow{2}{*}{\textbf{ISO}}
& \multicolumn{3}{c|}{PSNR estimation}
& \multicolumn{3}{c|}{SSIM estimation}
& \multicolumn{3}{c}{LPIPS estimation} \\
\cmidrule(lr){2-4}
\cmidrule(lr){5-7}
\cmidrule(lr){8-10}
& PLCC & SRCC & MAE
& PLCC & SRCC & MAE
& PLCC & SRCC & MAE \\
\midrule
100  & 0.935$\pm$0.004 & 0.932$\pm$0.005 & 1.861$\pm$0.026
     & 0.904$\pm$0.009 & 0.868$\pm$0.009 & 0.020$\pm$0.000
     & 0.669$\pm$0.025 & 0.794$\pm$0.014 & 0.031$\pm$0.001 \\

200  & 0.935$\pm$0.004 & 0.933$\pm$0.005 & 1.692$\pm$0.024
     & 0.909$\pm$0.009 & 0.865$\pm$0.010 & 0.020$\pm$0.000
     & 0.671$\pm$0.023 & 0.797$\pm$0.014 & 0.029$\pm$0.001 \\

400  & 0.935$\pm$0.004 & 0.933$\pm$0.005 & 1.437$\pm$0.023
     & 0.904$\pm$0.009 & 0.849$\pm$0.011 & 0.018$\pm$0.000
     & 0.704$\pm$0.024 & 0.793$\pm$0.014 & 0.025$\pm$0.001 \\

800  & 0.935$\pm$0.004 & 0.931$\pm$0.005 & 0.940$\pm$0.017
     & 0.898$\pm$0.010 & 0.863$\pm$0.012 & 0.016$\pm$0.000
     & 0.850$\pm$0.013 & 0.857$\pm$0.010 & 0.022$\pm$0.001 \\

1600 & 0.954$\pm$0.004 & 0.949$\pm$0.004 & 0.421$\pm$0.011
     & 0.958$\pm$0.005 & 0.957$\pm$0.005 & 0.011$\pm$0.000
     & 0.951$\pm$0.003 & 0.947$\pm$0.004 & 0.025$\pm$0.001 \\

3200 & 0.978$\pm$0.002 & 0.973$\pm$0.003 & 0.217$\pm$0.007
     & 0.983$\pm$0.002 & 0.987$\pm$0.001 & 0.009$\pm$0.000
     & 0.960$\pm$0.003 & 0.958$\pm$0.003 & 0.030$\pm$0.001 \\

6400 & 0.975$\pm$0.002 & 0.973$\pm$0.002 & 0.325$\pm$0.009
     & 0.987$\pm$0.001 & 0.991$\pm$0.001 & 0.013$\pm$0.000
     & 0.923$\pm$0.005 & 0.915$\pm$0.006 & 0.043$\pm$0.001 \\

\midrule
Average
     & 0.950$\pm$0.001 & 0.946$\pm$0.002 & 0.985$\pm$0.007
     & 0.935$\pm$0.003 & 0.911$\pm$0.003 & 0.015$\pm$0.000
     & 0.818$\pm$0.006 & 0.866$\pm$0.004 & 0.029$\pm$0.000 \\
\bottomrule
\end{tabular}
\end{adjustbox}
\end{table*}

\section{Impact of LoRA Layer Selection}
\label{sec:appendix_Impact_of_LoRA_Layer_Selection}

In most applications, LoRA fine-tuning is applied to the attention or
linear (MLP) layers.~\cite{hu2022lora, fomenko2024note,galim2024parameter,kwon2023datainf,lin2024tracking,zhang2025augmenting} Figure~\ref{fig:lora_layers} compares different
layer configurations in our setting. Although all variants achieve
similar metric prediction performance, the perceptual quality of the
estimated references differs. Since our SwinIR backbone includes
conditioning through Adaptive LayerNorm~\cite{peebles2023scalable}, restricting LoRA to the
attention $q,k,v$ projections leads to visible grid-like artifacts.
When extending adaptation to both the MLP and Adaptive LayerNorm
modules, these artifacts disappear, as shown in (e). Finally, adding
the VGG perceptual loss~\cite{ledig2017photo} further improves visual fidelity, yielding
results that are perceptually closer to the reference (f).

\subsection{Impact of LoRA rank}
\label{sec:appendix_Impact_of_LoRA_rank}
We further evaluated LoRA ranks of 8, 16, and 24. As shown in Table~\ref{tab:lora_ranks}, rank 16 achieves the best performance with slightly higher PLCC (0.955 vs. 0.952) and lower MAE (0.430 vs. 0.442) than rank 8. However, the improvement is marginal and within the confidence intervals. Increasing the rank to 24 does not lead to further gains and even degrades the performance (PLCC 0.948, MAE 0.532), likely due to over-parameterization. These results confirm that rank 8, which we adopted in the main paper, offers a good balance between computation efficiency and performance.
\begin{table*}[h]
  \centering
  \caption{\textbf{Impact of different LoRA ranks ($r=8,16,24$) on PSNR estimation evaluated using synthetic raw images.} We report the mean PLCC, SRCC, and MAE metrics with $\pm95$\% confidence intervals.}
  \label{tab:lora_ranks}
  \begin{adjustbox}{max width=\linewidth}
  \begin{tabular}{c|ccc|ccc|ccc}
    \toprule
    \multirow{2}{*}{\textbf{ISO}}
      & \multicolumn{3}{c|}{LoRA $r=8$}
      & \multicolumn{3}{c|}{LoRA $r=16$}
      & \multicolumn{3}{c}{LoRA $r=24$} \\
    \cmidrule(lr){2-4}\cmidrule(lr){5-7}\cmidrule(lr){8-10}
      & PLCC & SRCC & MAE & PLCC & SRCC & MAE & PLCC & SRCC & MAE \\
    \midrule
    100  & 0.920$\pm$0.010 & 0.956$\pm$0.003 & 0.590$\pm$0.024
         & 0.925$\pm$0.009 & 0.958$\pm$0.003 & 0.564$\pm$0.022
         & 0.923$\pm$0.008 & 0.959$\pm$0.003 & 0.605$\pm$0.024 \\
    200  & 0.929$\pm$0.010 & 0.960$\pm$0.003 & 0.561$\pm$0.021
         & 0.935$\pm$0.007 & 0.960$\pm$0.003 & 0.545$\pm$0.021
         & 0.932$\pm$0.007 & 0.962$\pm$0.003 & 0.571$\pm$0.021 \\
    400  & 0.946$\pm$0.007 & 0.962$\pm$0.003 & 0.514$\pm$0.017
         & 0.947$\pm$0.006 & 0.965$\pm$0.002 & 0.501$\pm$0.018
         & 0.945$\pm$0.006 & 0.966$\pm$0.002 & 0.523$\pm$0.019 \\
    800  & 0.958$\pm$0.003 & 0.966$\pm$0.002 & 0.463$\pm$0.013
         & 0.959$\pm$0.004 & 0.969$\pm$0.002 & 0.440$\pm$0.014
         & 0.956$\pm$0.003 & 0.970$\pm$0.002 & 0.476$\pm$0.015 \\
    1600 & 0.964$\pm$0.002 & 0.969$\pm$0.002 & 0.373$\pm$0.011
         & 0.966$\pm$0.003 & 0.971$\pm$0.002 & 0.360$\pm$0.011
         & 0.959$\pm$0.003 & 0.971$\pm$0.002 & 0.422$\pm$0.012 \\
    3200 & 0.968$\pm$0.002 & 0.976$\pm$0.002 & 0.304$\pm$0.010
         & 0.974$\pm$0.002 & 0.977$\pm$0.002 & 0.299$\pm$0.008
         & 0.958$\pm$0.004 & 0.972$\pm$0.002 & 0.446$\pm$0.012 \\
    6400 & 0.978$\pm$0.002 & 0.977$\pm$0.002 & 0.285$\pm$0.008
         & 0.979$\pm$0.002 & 0.980$\pm$0.002 & 0.301$\pm$0.008
         & 0.963$\pm$0.003 & 0.966$\pm$0.003 & 0.682$\pm$0.013 \\
    \midrule
    Average
         & 0.952$\pm$0.002 & 0.967$\pm$0.001 & 0.442$\pm$0.006
         & 0.955$\pm$0.002 & 0.969$\pm$0.001 & 0.430$\pm$0.006
         & 0.948$\pm$0.002 & 0.967$\pm$0.001 & 0.532$\pm$0.007 \\
    \bottomrule
  \end{tabular}
  \end{adjustbox}
\end{table*}


\end{document}